\documentclass[pdflatex,sn-mathphys-num]{sn-jnl}

\usepackage{graphicx}%
\usepackage{multirow}%
\usepackage{amsmath,amssymb,amsfonts}%
\usepackage{amsthm}%
\usepackage{mathrsfs}%
\usepackage[title]{appendix}%
\usepackage{xcolor}%
\usepackage{textcomp}%
\usepackage{manyfoot}%
\usepackage{amsmath,amsfonts}
\usepackage{algorithmic}
\usepackage[ruled,linesnumbered]{algorithm2e}
\usepackage{multirow}
\usepackage[caption=false,font=normalsize,labelfont=sf,textfont=sf]{subfig}
\usepackage{array}
\usepackage[section]{placeins}

\begin{document}

\title[Article Title]{Jellyfish: Zero-Shot Federated Unlearning Scheme with Knowledge Disentanglement}

\author{
Houzhe Wang, \
Institute of Information Engineering, \
Chinese Academy of Sciences, \
Beijing, China, \
\texttt{wanghouzhe@iie.ac.cn}
Xiaojie Zhu\footnote{Corresponding author}, \
King Abdullah University of Science and Technology, \
Thuwal, Kingdom of Saudi Arabia, \
\texttt{xiaojie.zhu@kaust.edu.sa}
Chi Chen, \
Institute of Information Engineering, \
Chinese Academy of Sciences, \
Beijing, China, \
\texttt{chenchi@iie.ac.cn}
}


\abstract{
With the increasing importance of data privacy and security, federated unlearning emerges as a new research field dedicated to ensuring that once specific data is deleted, federated learning models no longer retain or disclose related information.

In this paper, we propose a zero-shot federated unlearning scheme, named Jellyfish. It distinguishes itself from conventional federated unlearning frameworks in four key aspects: synthetic data generation, knowledge disentanglement, loss function design, and model repair. 
To preserve the privacy of forgotten data, we design a zero-shot unlearning mechanism that generates error-minimization noise as proxy data for the data to be forgotten. 
To maintain model utility, we first propose a knowledge disentanglement mechanism that regularises the output of the final convolutional layer by restricting the number of activated channels for the data to be forgotten and encouraging activation sparsity.  Next, we construct a comprehensive loss function that incorporates multiple components, including hard loss, confusion loss, distillation loss, model weight drift loss, gradient harmonization, and gradient masking, to effectively align the learning trajectories of the objectives of ``forgetting" and ``retaining".
Finally, we propose a zero-shot repair mechanism that leverages proxy data to restore model accuracy within acceptable bounds without accessing users' local data. 
To evaluate the performance of the proposed zero-shot federated unlearning scheme, we conducted comprehensive experiments across diverse settings. The results validate the effectiveness and robustness of the scheme.
}




\maketitle

\section{Introduction}\label{sec1}

Federated learning represents an advanced distributed machine learning paradigm that enables multiple clients to collaboratively train a shared global model without the need to share their local data \cite{konevcny2016federated} \cite{mcmahan2017communication}. 
This approach effectively overcomes a fundamental challenge in traditional machine learning by enabling model training without the need for centralized storage or processing of datasets.
Federated learning overcomes this limitation by enabling distributed training, allowing data holders to retain ownership and control of their data while collaboratively contributing to the development of a global model.
In this framework, participants only need to upload the updated parameters of their local models to a central server, which aggregates these updates to refine the global model. 
This approach effectively safeguards client data privacy by ensuring that raw data is retained locally.
\begin{figure}
    \centering
    \includegraphics[width=0.8\linewidth]{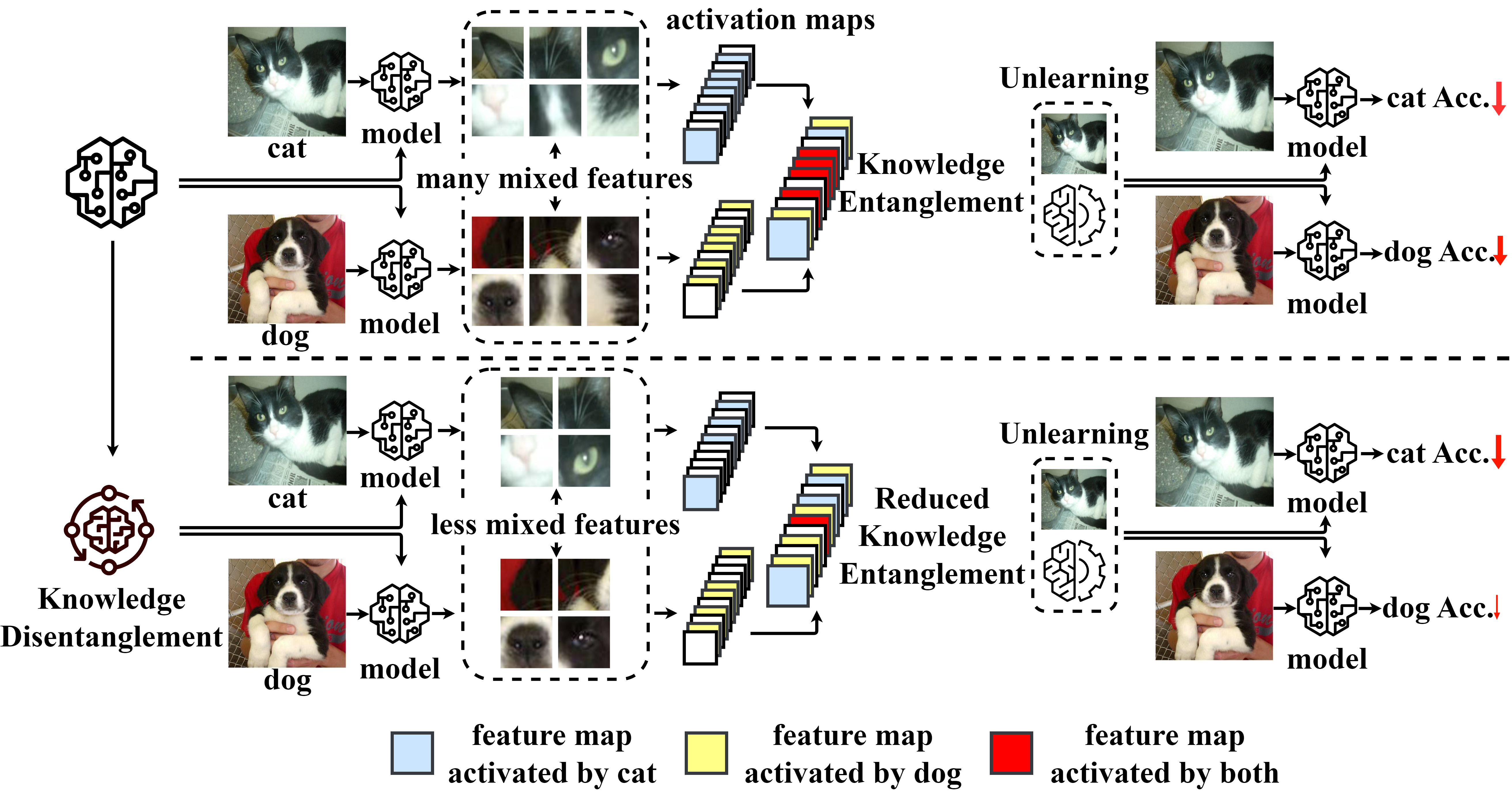}
    \caption{
    The motivation for knowledge disentanglement. The upper diagram illustrates the phenomenon of inter-class feature mixing during the feature extraction process in conventional models. This mixing leads to the simultaneous activation of features for cats and dogs in the activation maps (as shown in the red areas), affecting the accurate representation of single-class features.
    The lower diagram demonstrates the effect after applying knowledge disentanglement, where the model learns to represent features with less mixing. The activation maps show clearer class specificity (blue is primarily related to cats, and yellow to dogs). This disentanglement reduces activation overlap between different classes, enhancing the model's interpretability.}
    \label{disentanglefigure}
\end{figure}

However, within the federated learning framework, users may request the removal of their contribution from the trained global model. Moreover, recent regulations such as the European Union's General Data Protection Regulation (GDPR) \cite{regulation2018general} and the California Consumer Privacy Act (CCPA) \cite{pardau2018california} empower individuals with the right to demand the deletion of their private data from any part of the system within a reasonable time frame. Furthermore, even if the original data were never shared, the global machine learning model could still glean information about the clients \cite{nasr2018comprehensive} \cite{song2020analyzing}. The predictions made by the global model could potentially leak client information \cite{salem2018ml}\cite{bagdasaryan2020backdoor}. Therefore, a compelling need arises for a method to effectively eliminate a client's contribution from the trained global model.

Two approaches exist to ensure that a global model forgets the contributions of a specific client. The first involves retraining the model from scratch after excluding the data of the target user \cite{guo2019certified}. The second approach directly removes the user information from the trained model's parameters while preserving its overall utility\cite{jeong2024sok}. However, retraining becomes impractical when dealing with large datasets and complex models due to its significant time and energy demands, high computational costs, and scalability challenges. Consequently, there is a growing interest in developing cost-effective federated machine unlearning algorithms that efficiently mitigate the influence of deleted data on trained models.

To efficiently mitigate the impact of deleted data on trained models, we identified the critical importance of knowledge disentanglement.
 As shown in Figure \ref{disentanglefigure},
we introduce the motivation for the disentanglement of knowledge. 
Conventional models extract mixed features from many classes
\cite{zhang2018interpretable}, known as knowledge entanglement \cite{lin2023erm}, hindering interpretability and class-specific feature representation. 
We aim to reduce mixed features through knowledge disentanglement, ensuring the extraction of features that correspond to only one (or a few) classes, making them class-specific.
This enhances interpretability and strengthens the representation of class-related features. It ensures that the removal of specific data does not negatively impact the performance or accuracy of other unrelated data in the model.
This process is essential to ensure that the removal of specific data does not negatively impact the performance or accuracy of other unrelated data in the model.
However, disentangling knowledge is a challenging task, as representations from different classes often become interwined during the training process. 
This entanglement can lead to over-forgetfulness during the unlearning process, where removing data from one class inadvertently degrades the accuracy of other unrelated classes.
Lin \textit{et al.}  highlighted that the most representative knowledge learned by machine learning models is the features extracted by the convolutional layers \cite{lin2023erm}. 
However, the features of different classes are intertwined, making it challenging to transfer knowledge of the target class without affecting other classes. 
Zhu \textit{et al.} attempted to address the scenario in which there is a mismatch between the target concept label domain and the label domain of the data to be forgotten \cite{zhu2024decoupling}. This misalignment can lead to two key issues: over-forgetfulness or insufficient forgetting. 
 Liang \textit{et al.} pointed out that filters in CNNs typically extract features mixed with various semantic concepts, including objects, parts, scenes, textures, materials, and color categories \cite{liang2020training}. 
Therefore, reducing entanglement is crucial for humans to interpret the concepts of filters.

In addition, in conventional federated unlearning, users are required to submit the target data to be deleted to the server.  This process often involves the use of private information, which raises concerns about maintaining the privacy and security of the user's sensitive data during the unlearning process.
Tarun \textit{et al.} \cite{tarun2023fast} and Chundawat \textit{et al.} \cite{chundawat2023zero}  observed that in specific cases, machine learning models are trained using sensitive data such as facial images and personal medical information. Given the highly sensitive nature of these data and the constraints imposed by data protection regulations such as GDPR \cite{regulation2018general} and CCPA \cite{pardau2018california}, it may not be feasible to use the original data to perform the unlearning process, even when asked to forget. 
Lastly, the unlearning process often involves two conflicting optimization objectives: first, ensuring that the model forgets the specified data, and second, preserving the model's knowledge of the remaining data. Balancing these objectives is crucial to achieving efficient and effective unlearning without degrading the performance of non-removed data. 

Several studies \cite{foster2024fast,golatkar2020eternal,golatkar2020forgetting,liu2023muter,mehta2022deep}  proposed perturbing model weights based on computational estimates of the impact of forgotten data. This technique selectively adjusts the parameters most influenced by the removal of specific data, allowing the model to effectively forget the targeted information while preserving the integrity of the remaining learned knowledge. 
While these methods can effectively diminish the model's capabilities on forgetting dataset \( D_f \), 
careful calibration is crucial to avoid over-forgetfulness. Excessive forgetting can lead to a significant decline in the model's performance on the remaining dataset \( D_r \) after the removal of the targeted data, potentially resulting in a collapse of its overall effectiveness. 

 To improve the model's ability to retain knowledge of the non-targeted data, recent studies \cite{chourasia2023forget,chundawat2023can,fan2023salun,graves2021amnesiac,kim2022efficient,tarun2023fast} have incorporated training or distillation using \( D_r \) into the forgetting process. 
Unlike retraining from scratch, these methods selectively focus on reinforcing the model's knowledge of the non-targeted data. 
By incorporating direct training signals from \( D_r \), these methods guide the model to relearn essential information that may be inadvertently lost during the unlearning process, without the need for full retraining. 

Particularly, recent works \cite{alam2024get,xia2023fedme,dinsdale2022fedharmony,wang2024goldfish,wang2023bfu} often incorporated multiple objectives into a single loss function, aiming for multi-objective optimization by minimizing the combined loss. However, since different components of the loss function correspond to distinct optimization paths, this results in optimization challenges \cite{jeong2024sok}.

To address the aforementioned challenges, we propose a novel federated unlearning paradigm, named Jellyfish. 
{The key innovations of our work are highlighted as follows:
\begin{enumerate}
    \item Knowledge disentanglement: To mitigate knowledge entanglement, we introduce a knowledge disentanglement mechanism that regularizes the output of the final convolutional layer by limiting the number of activation channels for the data to be forgotten and promoting sparsity in activations.
    \item Zero-shot Unlearning and Repair Mechanism: To safeguard the privacy of forgotten data, we design a zero-shot unlearning mechanism through generating error-minimization noise as proxy data for the forgotten data. Furthermore, to maintain model performance, we propose a zero-shot repair mechanism that leverages proxy data to restore accuracy within acceptable bounds without accessing users' local data.
    \item Multi-Objective Optimization with Gradient Harmonization and Masking: To reconcile conflicting optimization objectives, we design a comprehensive loss function that integrates hard loss, confusion loss, distillation loss, model weight drift loss, gradient harmonization, and gradient masking, effectively aligning the learning trajectories of the ``forgetting" and ``retaining" tasks.
\end{enumerate}

Finally, we conduct comprehensive experiments, and the results illustrate the effectiveness and robustness of the proposed scheme. 
}

\begin{algorithm}[h]
    \caption{Jellyfish: Zero-Shot Federated Unlearning Scheme}
    \label{alg:ZeroShotUnlearning}
    \SetAlgoLined
    \SetKwInOut{Input}{Input}
    \SetKwInOut{Output}{Output}
    \SetKwProg{Procedure}{Procedure}{:}{}

    \Input{initialized model parameter $\omega^0$, global model parameter $\omega^t$, reference model parameter $\omega^{\text{ref}}$, disentangling learning rate $\mu_{\text{dis}}$, unlearning learning rate $\mu_{\text{un}}$, repair learning rate $\mu_{\text{re}}$, disentangling epoch number $E_{\text{dis}}$, unlearning epoch number $E_{\text{un}}$, performance repair epoch number $E_{\text{re}}$ (optional), user deletion request $D_f$, user remaining dataset $D_r$ (optional), disentangling threshold $\alpha$}
    \Output{unlearned global model}

    \Procedure{Federated Unlearning}{}{
        
        $D_f, D_r \leftarrow Error$-$Minimization$ $Noise(D_f,D_r)$\;
        $\omega^{\text{ref}} \leftarrow \omega^t$\;
        \vspace{0.5cm}
        \textbf{Disentangle}:\\
        \For{epoch $= 1,2,3,\ldots, E_{\text{dis}}$}{
            \ForEach{batch in $D_f$}{
                $\mathcal{L}_{\text{disentangle}} \leftarrow \text{calculate Disentangle loss}(D_f, \alpha)$ using equation \ref{disentangle loss}\;
                $\omega^t \leftarrow \omega^t - \mu_{\text{dis}} \nabla_{\omega} \mathcal{L}_{\text{disentangle}}$\;
            }
        }
        \textbf{Unlearn}: \\
        \For{epoch $= 1,2,3,\ldots, E_{\text{un}}$}{
            \ForEach{batch in $D_f$}{
                $\mathcal{L}_{\text{unlearn}} \leftarrow \text{calculate unlearn loss}(batch, \omega^{t}, \omega^0)$ using equation \ref{total unlearn loss}\;
                $\mathcal{L}_{\text{drift}} \leftarrow \text{calculate drift loss}(\omega^{\text{ref}}, \omega^t)$ using equation \ref{drift loss}\;
                $g_f \leftarrow \nabla_{\omega} \mathcal{L}_{\text{unlearn}}$\;
                $g_r \leftarrow \nabla_{\omega} \mathcal{L}_{\text{drift}}$\;
                $g_r' \leftarrow \text{Gradient Masking}(g_r, g_f, \pi)$ using equation \ref{masking}\;
                $G \leftarrow \text{Gradient Harmonization}(g_f, g_r')$\;
                $\omega^t \leftarrow \omega^t - \mu_{\text{un}} G$ using equation \ref{gf'}\;
            }
        }
            \textbf{Repair(optional):}\\ 
            $D_r \leftarrow N_r\_list$\;
            \For{epoch $= 1,2,3,\ldots, E_{\text{re}}$}{
                \ForEach{batch in $D_r$}{
                    $x_r, y_r \leftarrow D_r$\;
                    $\mathcal{L}_{\text{repair}} \leftarrow \text{MSE}(M(x_r), y_r)$\;
                    $\omega^t \leftarrow \omega^t - \mu_{\text{re}} \nabla_{\omega^t} \mathcal{L}_{\text{repair}}$\;
                }
            }
        
        \Return $\omega^t$\;
    }

\end{algorithm}

\section{Related Work}

In this paper, we propose a federated unlearning scheme based on knowledge disentanglement.
To properly position our contribution, we first review relevant prior work and then highlight the key differences between our approach and existing methods.  

{\subsection{Exact Unlearning}
Exact unlearning is based on the method of retraining from scratch. It mainly improves the training process of the model, so that when the model needs to forget data, it can reduce the computational and time costs of model training.

Bourtoule \textit{et al.} \cite{bourtoule2021machine} introduced SISA training as an approach to alleviate the computational costs associated with forgetting.
Building upon the SISA framework, several related studies have been proposed.
The random forest algorithm improves the performance of the model by building multiple decision trees and summarizing their predictions \cite{brophy2021machine}. 
In terms of unlearning, each tree corresponds to a slice in the SISA framework, 
trained independently and isolated from the influence of data points, 
so that when it is necessary to forget specific data points, only those trees that include the data point need to be retrained.
DC-k-means \cite{ginart2019making}, as an extension of k-means, uses a tree-based hierarchical clustering method to achieve exact unlearning. It randomly divides the data into multiple subsets and trains a k-means model on each subset, finally constructing the final clustering result by merging these models.
KNOT \cite{su2023asynchronous} uses the SISA framework to implement client-level asynchronous exact unlearning. Through cluster aggregation, clients are grouped, and the server aggregates the model within the cluster, while each cluster trains independently. Data deletion requests only trigger retraining of clients in the same cluster.

Additionally, there are researchers exploring the relationship between the model and data, as well as efforts focused on enhancing training efficiency.
\cite{cao2015towards} draws inspiration from statistical query learning and designs an intermediate layer called ``Summation", which serves as a buffer between the machine learning algorithm and the training data, making the algorithm learn through summarized statistical information rather than original data.
Liu et al. 
\cite{liu2022right} uses the first-order Taylor expansion approximation technique to customize a rapid retraining algorithm based on diagonal experience FIM.

\subsection{Approximate Unlearning}
Approximate unlearning aims to minimize the impact of data that needs to be deleted or forgotten to an acceptable level while also achieving an efficient unlearning process.

Compared to exact unlearning techniques, approximate unlearning offers several advantages, including better computational efficiency, lower storage costs, and greater flexibility.
In terms of computational efficiency, approximate unlearning methods reduce computational costs by minimizing rather than completely deleting the impact of data, as opposed to exact unlearning methods that require retraining with the remaining data \cite{guo2019certified}. For instance, the method proposed in \cite{guo2019certified} adjusts model parameters to reduce the influence of specific data, thus reducing computational intensity compared to exact unlearning.
Regarding storage overhead, approximate unlearning methods, such as those presented by Sekhari \textit{et al.} in \cite{sekhari2021remember}, store only necessary statistical information of the data, thereby significantly reducing storage costs.
In terms of flexibility, approximate unlearning methods are highly adaptable, as they typically do not rely on specific learning models or data structures, allowing for broader application to a variety of learning algorithms \cite{cao2015towards}. This flexibility is related to the trade-off between completeness and efficiency made by approximate unlearning, allowing for adaptation to new data and tasks by accelerating the unlearning process and reducing costs while maintaining model performance. For example, 
Zhang \textit{et al.} \cite{zhang2023fedrecovery} eliminate client influence by extracting the weighted sum of gradient residuals from the global model and introducing Gaussian noise. This process is designed to achieve statistical indistinguishability between the unlearned and retrained models.

Liu \textit{et al.} \cite{liu2021federaser} reconstruct the forgotten model using parameter updates stored on the server, introducing a novel calibration method to adjust client updates. This innovative approach aims to enhance the speed of unlearning while preserving model performance.
{
Meng \textit{et al.} \cite{meng2025survey} analyze the privacy threats in semantic communication across its training, encoding, and transmission stages, demonstrating how defense techniques can enhance the privacy assurance of federated unlearning. This comprehensive survey provides a valuable reference for constructing an end-to-end secure data lifecycle management framework.}
Additionally, Baumhauer \textit{et al.} \cite{baumhauer2022machine} and Thudi \textit{et al.} \cite{thudi2022necessity} emphasize the pursuit of higher efficiency in machine unlearning by relaxing the requirements for both effectiveness and provability.
Izzo \textit{et al.} \cite{izzo2021approximate}, Neel \textit{et al.} \cite{neel2021descent}, and Wu \textit{et al.} \cite{wu2020deltagrad} explore techniques for the server to effectively approximate gradients during the unlearning process by leveraging historical gradients and model weights.
Chourasia \textit{et al.} \cite{chourasia2023forget} enhances model robustness in addressing data deletion.
Halimi \textit{et al.} \cite{halimi2022federated} and Wu \textit{et al.} \cite{wu2022federated} employed a gradient-based approach to forget data, using the gradient information
from the forgetting set.
Wu et al. \cite{wu2022federated1} and Zhu et al. \cite{zhu2023heterogeneous}
explored knowledge distillation to selectively remove data
from models, enhancing the unlearning process in federated
learning environments.
Wang \textit{et al.} \cite{wang2024goldfish} propose Goldfish, an efficient federated unlearning framework, consisting of four modules: basic model, loss function, optimization, and extension. Each module is crafted to enhance the practicality of the framework.

\subsection{Zero-shot Unlearning}

In recent years, a novel paradigm of machine unlearning, called zero-shot unlearning, has emerged. This approach aims to achieve unlearning without requiring access to the original data \cite{ghazal2024zero}.

Chundawat \textit{et al.} \cite{chundawat2023zero} were the pioneers in proposing methods for the zero-shot unlearning setup, where neither the retained data \( D_r \) nor the data requested for forgetting \( D_f \) are accessible.
UNSIR \cite{tarun2023fast} introduced the concept of zero-glance unlearning, a technique that allows data forgetting in a zero-observation privacy setting, where the model does not have visibility into the data categories that need to be forgotten.
EMMN \cite{chundawat2023zero} is a zero-shot machine unlearning technique that extends UNSIR \cite{tarun2023fast} by maximizing the loss related to the forgotten categories to obtain noise for data forgetting and minimizing the loss of the remaining data to obtain noise for repair.
{Fan \textit{et al.} \cite{fan2025generative} demonstrate that diffusion models can effectively fit complex data distributions through a forward noising and reverse denoising process, thereby generating synthetic samples that preserve the statistical properties of the original data.}
Gated Knowledge Transfer (GKT) \cite{chundawat2023zero} is a zero-shot machine unlearning method aimed at achieving data forgetting through a knowledge distillation strategy.
Zero-shot unlearning using Lipschitz Regularization (JiT) \cite{foster2024zero} is also a zero-shot unlearning technique that leverages Lipschitz continuity to minimize the model's output sensitivity to input perturbations \cite{yoshida2017spectral}, thereby achieving the forgetting of specific data points while maintaining the overall performance of the model.}

\subsection{Knowledge disentanglement}
In the domain of knowledge disentanglement, recent research has primarily focused on effectively removing the influence of specific data points from machine learning models, with a strong emphasis on preserving data privacy. 

Zhu \textit{et al.} \cite{zhu2024decoupling} introduced a framework known as TARget-aware Forgetting (TARF), which is designed to achieve knowledge disentanglement by differentiating between concepts. 
The TARF framework isolates target concepts through annealed gradient ascent on data to be forgotten, combined with selective gradient descent on the remaining data.
This approach enables more precise forgetting of specific concepts while preserving model performance. 
Zhu \textit{et al.}'s work offers a novel perspective in the field of machine unlearning, particularly for complex forgetting scenarios that require a careful balance between unlearning effectiveness and model performance retention. 
{While effective in isolating target concepts, its requirement for direct access to the original forgetting data poses a privacy risk in federated settings, making it less suitable for strict privacy regulations like GDPR}.

Liang \textit{et al.} \cite{liang2020training} aimed to enhance the interpretability of Convolutional Neural Networks (CNNs) by introducing a learnable sparse Class-Specific Gate (CSG) structure. This design encourages the development of class-specific filters, where each filter responds only to one or a few classes.
Their approach effectively reduces the filter-class entanglement, \textit{i.e.}, the complex correspondence between filters and classes. Moreover, it demonstrates practical advantages in tasks such as object localization and adversarial sample detection.
{However, their method is primarily designed for model interpretability during training, not for the unlearning process post-training.}

Shen \textit{et al.} \cite{lin2023erm} introduced the ERM-KTP method, which defines knowledge disentanglement from a knowledge perspective and proposes a knowledge-level machine decoupling approach.
During training, an Entanglement Reduction Mask (ERM) is employed to reduce the entanglement of knowledge points across different classes. Upon receiving an unlearning request, the Knowledge Transfer and Prohibition (KTP) mechanism transfers the knowledge of non-target data points from the original model to a decoupled model, while explicitly prohibiting the transfer of target knowledge. This approach not only improves the interpretability of the disentanglement process but also achieves strong performance in terms of efficiency, fidelity, and scalability.
{Nevertheless, similar to \cite{zhu2024decoupling}, it operates under the assumption that the data to be forgotten is accessible, which conflicts with the zero-shot principle we aim to uphold.}



{
The proposed Jellyfish scheme distinguishes itself from existing federated unlearning approaches through a systematic integration of adapted and novel components. While it incorporates effective concepts from prior work---such as the use of noise for inducing randomness \cite{chundawat2023can} and error-minimization-based data proxying \cite{chundawat2023zero}---its core contributions are uniquely tailored to the federated setting. These contributions are fourfold:

First, the framework establishes a complete zero-shot unlearning pipeline. Unlike methods that require access to the original forgotten data \cite{zhu2024decoupling,lin2023erm}, Jellyfish operates entirely using proxy data, ensuring that no raw data is accessed or transmitted during the unlearning process.

Second, we introduce a novel knowledge disentanglement mechanism that is formally integrated as an optimization objective within the training process. This approach explicitly reduces feature entanglement across classes by sparsifying activation channels related to the data to be forgotten, thereby enhancing both the precision of unlearning and model interpretability.

Third, the loss function incorporates a new form of confusion loss and combines multiple objectives, including hard loss, distillation loss, model weight drift loss, gradient harmonization, and gradient masking, into a unified optimization framework.

Fourth, the proposed zero-shot repair mechanism enables model performance recovery without relying on any original remaining data. By utilizing proxy data for repair, Jellyfish maintains full compliance with data privacy constraints while preserving model utility.

In summary, Jellyfish advances the state of the art by introducing federated-specific innovations in disentanglement, loss design, and end-to-end zero-shot capability, setting it apart from existing partial or non-federated alternatives.}

\section{Preliminaries}
In this section, we provide background on federated learning and federated unlearning.
Additionally, the frequently used notations are summarized in Table \ref{tab:notations}.

\begin{table}[]
    \centering
    \renewcommand\arraystretch{1}
    \caption{Summary of Notations}
    \label{tab:notations}
    \begin{tabular}{|c|l|}
    \hline
    Notations            & Description                            \\ \hline
    $N_{\text{client}}$  & Number of clients                      \\
    $c$                  & Client index                           \\
    $D$                  & Complete dataset                       \\
    $D_f$                & forgetting dataset                     \\
    $D_r$                & remaining dataset                      \\
    $D_c$                & Local dataset                          \\
    $A(\cdot)$           & Learning algorithm                     \\
    $U(\cdot)$           & Unlearning algorithm                   \\
    $\omega^0$           & initialized global model               \\
    $\omega^t$           & global model at $t$-th round           \\
    $\omega_c^t$         & Client c’s local model at $t$-th round \\
    $\omega^{\text{un}}$ & Unlearned model                        \\
    $\omega^*$           & Retrained model                        \\ \hline
    \end{tabular}
\end{table}

\subsection{Federated Learning}
Federated Learning (FL) \cite{mcmahan2017communication} is a distributed machine learning paradigm that has recently garnered significant attention. It allows individuals to collaborate in training a global machine learning model without the need to share their private training data with others. FL typically consists of a server and \( N_{\text{clients}} \) clients. During the FL training process, at the initial stage, each client \( c \) initializes its local client model \( \omega_c^0 \) using the initialized global model \( \omega^0 \). 
In the subsequent rounds, each client \( c \) conducts local training using the current round \( t \) global model \( \omega^t \) and its local training data \( D_c \), with the learning algorithm denoted as \( A(\omega_c^t, D_c) = \omega_c^{t+1} \), and then sends its local model update \( \omega_c^{t+1} \) to the server. After receiving model updates from all clients, the server utilizes a specific aggregation rule to combine the received model updates and update the global model. The updated global model \( \omega^{t+1} \) is then distributed to all clients for the next round of training.

For instance, the FedAvg \cite{mcmahan2017communication} aggregation rule calculates the average of model updates to obtain the global model, which is used in non-adversarial scenarios. One advantage of FL over centralized learning is that clients no longer need to send their private training data to the server. The model aggregation scheme of FedAvg is shown as Equation \ref{fedavg_aggregation}:
\begin{equation}
    \omega^{t+1} = \frac{1}{N_{\text{clients}}} \sum_{c=1}^{N_{\text{clients}}} \omega_c^{t+1}
    \label{fedavg_aggregation}
\end{equation}
where $N_{\text{clients}}$ is the total number of clients.

\subsection{Federated Unlearning}

Federated Unlearning (FU) has emerged as a critical strategy within FL, enabling the elimination of the influence of specific knowledge, whether individual data points, features, or broader data concepts, from a pre-trained FL model without requiring complete retraining from scratch. 
This capability is particularly crucial in federated environments, where data privacy and computational efficiency are paramount. The subset of knowledge designated for removal is referred to as the forgetting set. The primary objective of FU is to efficiently update the pre-trained FL model such that its performance closely approximates that of a model retrained from scratch, excluding the forgetting set, while preserving the integrity of the remaining knowledge.

Upon receiving user deletion requests, this entire dataset $D$ is partitioned into two disjoint subsets: the forgetting set \( D_f \) and the retention set \( D_r \), such that \( D_f \cap D_r = \emptyset \) and \( D_f \cup D_r = D \). 
Upon receiving unlearning requests, the server applies an unlearning algorithm \( U(\omega^t, D_f, D_r) = \omega^{un} \), where \( \omega^{un} \) denotes the model parameters after unlearning. Note that $D_f$ and $D_r$ are optional depending on the specific unlearning scenarios. As shown in the following equation, the optimization goal is to ensure that \( \omega^{un} \) approximates the parameters \( \omega^* \) of a model retrained from scratch on $D_r$, thus maintaining performance while removing the influence of $D_f$. 
\begin{equation}
    \min_{\omega^{un}} Dis(\omega^{un}, \omega^*, D) \quad \text{where} \quad \omega^* = A(\omega^0, D_r)
\end{equation}
Here, \( Dis \) denotes an evaluation function used to quantify the difference between two models (\textit{e.g.}, based on loss, accuracy on the same dataset, or other metrics).
The task of federated unlearning inherently involves both ``forgetting" and ``remembering", making it conceptually similar to multi-task learning \cite{jeong2024sok}. 
In line with this perspective, the loss function in Goldfish \cite{wang2024goldfish} is crafted to 
address multiple objectives: (1) minimizing the hard loss between model predictions and ground-truth labels on the remaining dataset \(D_r\), (2) mitigating the bias in predictions on the removed dataset \(D_f\), and (3) assessing the confidence of the model's predictions on $D_f$. 
In this paper, we adopt a similar principle and further enhance the loss function design by incorporating additional components aimed at improving overall performance and robustness.

\section{Proposed Federated Unlearning Scheme: Jellyfish}
\begin{figure}
    \centering
    \includegraphics[width=1\linewidth]{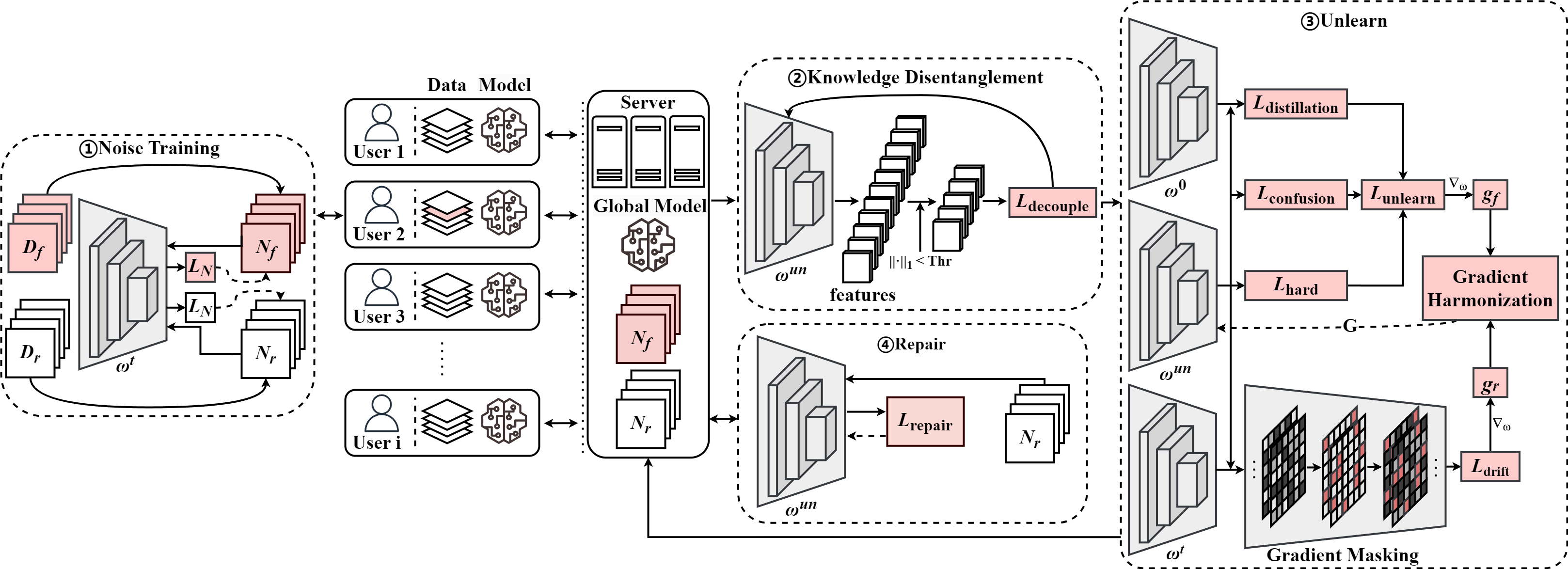}
    \caption{Proposed Jellyfish Scheme:
    \textcircled{1} Noise Training: The user requests data deletion locally and trains proxy data   $N_f$  to replace the deleted data.
    \textcircled{2} Knowledge Disentanglement: When the server receives the data, it first disentangles the data to reduce knowledge entanglement between categories.
    \textcircled{3} Unlearn: The model is guided to forget by using an improved loss function and mechanisms like gradient harmonization.
    \textcircled{4} Repair: If the model's accuracy drops after unlearning, the proxy data of the remaining data  $N_r$ is used to restore the model's performance.
    }
    \label{Unlearning Method}
\end{figure}

In this section, we first provide an overview of the proposed federated unlearning scheme: Jellyfish. 
We then introduce the approach for generating proxy data to approximate the forgotten data. 
Following this, we propose the knowledge disentanglement technique tailored for forgotten data, and detail the novel construction of the loss function. Finally, we conclude with a description of the proposed model repair procedure. 

\subsection{Overview}
In this section, we outline the proposed zero-shot federated unlearning scheme. 
As shown in Figure \ref{Unlearning Method}, when a user $i$ issues a forgetting request for a dataset $ D_f^i$, the server avoids direct data sharing by employing an error-minimization noise approach to generate proxy data that approximates the forgotten samples.
Next, we perform knowledge disentanglement on the proxy data, constraining its influence within specific channels of the model and minimizing its association with unrelated categories. 
Subsequently, we design a comprehensive loss function to guide the model through the forgetting process. This loss function consists of six components: hard loss, confusion loss, distillation loss, model weight drift loss, gradient harmonization, and gradient masking. 
The hard loss explicitly targets the information to be forgotten, while the confusion loss and distillation loss focus on redistributing the class-related information of the forgotten data.
To mitigate the performance degradation resulting from the forgetting process, the model weight drift loss is introduced to compute the gradient updates that help preserve the model's original performance. 
Additionally, a gradient harmonization mechanism is employed to resolve conflicts between gradient directions arising from the competing objectives of forgetting and retention. 
Finally, gradient masking is applied to suppress the retention of knowledge associated with forgotten data, ensuring a more precise and effective unlearning process.
Once the server completes the unlearning process, it redistributes the unlearned global model to all users. 
If some users observe a significant decline in model performance, they can generate error-minimization noise to construct proxy data that approximates the removed dataset $D_r$. These proxy data are then sent to the server to initiate a repair process, which helps restore the model’s performance to an acceptable level.

\subsection{Error-Minimization Noise}

\begin{algorithm}[ht]
    \caption{Error-Minimization Noise}
    \label{alg:ErrorMinimizationNoise}
    \SetAlgoLined
    \SetKwInOut{Input}{Input}
    \SetKwInOut{Output}{Output}
    \SetKwProg{Procedure}{Procedure}{:}{}
    \Input{global model $M$, training noise learning rate $\mu_{\text{no}}$, training noise epoch number $E_{\text{no}}$}
    \Output{$N_f$, $N_r$ (optional)}
    \vspace{0.2cm}
    \Procedure{Error-Minimization Noise($N_f, N_r$)}{}{
    $N_f\_list \leftarrow [\ ]$\;
    \ForEach{Class\_data in $D_f$}{
        \For{epoch $= 1,2,3,\ldots, E_{\text{no}}$}{
            Get batch size $B$, channel dimension $C$, data size $H$, $W$ from Class\_data\;
            Initialize noise matrix $N_f \in \mathbb{R}^{B \times C \times H \times W}$\;
            $N_f \leftarrow \text{TrainNoiseData}(Class\_data, N_f, \mu_{\text{no}})$\;
            $N_f\_list.append(N_f)$\;
        }
    }
    \vspace{0.2cm}
    \textbf{optional:} \\
    $N_r\_list \leftarrow [\ ]$\;
    \ForEach{Class\_data in $D_r$}{
        \For{epoch $= 1,2,3,\ldots, E_{\text{no}}$}{
            Get batch size $B$, channel dimension $C$, data size $H$, $W$ from Class\_data\;
            Initialize noise matrix $N_r \in \mathbb{R}^{B \times C \times H \times W}$\;
            $N_r \leftarrow \text{TrainNoiseData}(Class\_data, N_r, \mu_{\text{no}})$\;
            $N_r\_list.append(N_r)$\;
        }
    }
    
    \Return $N_f\_list, N_r\_list$\;
    }
    \vspace{0.2cm}
    \Procedure{TrainNoiseData($Class\_data, N, \mu_{\text{no}}$)}{}{
        \ForEach{batch in Class\_data}{
            imgs, labels $\leftarrow$ Class\_data\;
            $\mathcal{L}_N \leftarrow \text{calculate noise loss using equation \ref{Nfnoise loss}}$\;
            $N \leftarrow N - \mu_{\text{no}} \nabla_N \mathcal{L}_N$\;
        }
        \Return $N$\;
    }
\end{algorithm}

In federated learning, users collaboratively train a global model without sharing their raw data. However, when a user submits a data deletion request, the user must provide the specific data to be removed, which poses a significant risk of data privacy leakage. 
This concern is especially critical in contexts involving sensitive information, such as facial images or personal medical records. 
Furthermore, strict data protection regulations, such as the GDPR \cite{regulation2018general} and CCPA\cite{pardau2018california}, impose strict time constraints on processing user data deletion requests. As a result, even the data necessary for executing the unlearning process may be restricted from further use \cite{tarun2023fast}.    

Our objective is to enable users to request immediate deletion of their data and ensure their complete removal from the trained model. After the model weights are updated, the model should no longer retain any information related to the deleted data. 


To preserve data privacy, we introduce zero-shot unlearning techniques into the federated unlearning framework. 
Unlike conventional federated unlearning methods, which typically require users to submit the data they wish to delete as part of a forgetting request processed by the central server,  our approach eliminates the need for direct access to the original data, thus enhancing data privacy and compliance with strict regulatory requirements. 
To mitigate the potential data privacy risks associated with transmitting deletion requests to the server, inspired by \cite{chundawat2023zero, tarun2023fast}, we design an algorithm that generates error-minimization noise as a proxy for \( D_f \). 
{Intuitively, this noise is not derived from the statistical properties of the original data but is generated through an adversarial optimization process. We start with a random noise matrix and iteratively adjust it by minimizing the prediction error of the current model on the target class (as defined in Equation \ref{Nfnoise loss}). The core idea is to create synthetic data that the model confidently recognizes as the class to be forgotten, without this data containing any recognizable features from the actual private dataset. Since only the optimized noise, which is devoid of any genuine data patterns, is shared with the server, the privacy of the forgotten data is effectively preserved.}
Specifically, when a user $i$ submits a deletion request $D_f^i$, which consists of $n_B$ batches, the user first locally trains a noise matrix \( N_f^i(j) \in \mathbb{R}^{B \times C \times H \times W} \), where \( B \) represents the batch size, \( C \) is the number of channels,   and $H$ and $W$  represent the dimensions of the actual sample size. 
{ $N_f^i(j)$ is initialized by sampling from a standard Gaussian distribution, i.e., $N_f^i(j) \sim \mathcal{N}(0, 1)$, which provides a neutral starting point for optimization.}
The index \( j \) corresponds to the batch number, where \( j \in [1, n_B] \). The noise matrix $N_f^i$ serves as a proxy for \( D_f^i \), and is transmitted to the server for use in the subsequent unlearning process.  The  noise matrix is optimized locally by  minimizing the loss function in Equation \ref{Nfnoise loss}:
\begin{equation}
\label{Nfnoise loss}
    \mathcal{L}_{N_f^i} = \frac{1}{n_B} \sum_{j=1}^{n_B} -y_f^i(j) \log M(N_f^i(j))
\end{equation}
where \( y_f^i(j) \) represents the class label of the data to be forgotten, \( n_B \) denotes the number of batches in the noise matrix, and \( M(N_f^i(j)) \) refers to the predicted probability distribution generated by inputting the $j$-th batch of noise $N_f^i(j)$ into the classifier model being unlearned.

In scenarios  
where multiple users \( i = 1, 2, \ldots, n_f \) simultaneously submit deletion requests \( D_f^i \), 
each user independently trains a noise matrix \( N_f^i \) locally. These noise matrices serve as proxies for their respective deletion requests and are then 
transmitted to the central server for further unlearning processing.
{ The local training process for each user (lines 4-9 in Algorithm \ref{alg:ErrorMinimizationNoise}) continues for a fixed number of epochs $E_{no}$, which serves as a straightforward convergence criterion to ensure computational efficiency and predictability. While users train their noise matrices independently, potential inconsistencies in noise quality across users are mitigated by the server-side aggregation defined in Equation \ref{union Nf loss}. This aggregation averages the contributions from all users, inherently reducing the variance introduced by any single low-quality noise matrix. Furthermore, all users adhere to the same hyperparameters (e.g., learning rate $\mu_{no}$) during local noise training, promoting a baseline level of consistency in the optimization process across different clients.}
On the server side, the received noise matrices \( \{N_f^1, N_f^2, \ldots, N_f^{n_f}\} \) are aggregated into a unified noise matrix set,  denoted as \( N_f \). This combined matrix \( N_f \) effectively simulates the collective influence of all the data targeted for deletion. The optimization objective for \( N_f \) is defined as the synthesis of the local optimization goals from each user. Specifically, this objective is expressed in Equation \ref{union Nf loss}:

\begin{equation}
    \mathcal{L}_{N_f} = \frac{1}{n_f} \sum_{i=1}^{ n_f} \frac{1}{n_B^i}\sum_{j=1}^{n_B^i} -y_f^i(j) \log M(N_f^i(j))
    \label{union Nf loss}
\end{equation}
where \(  n_f \) is the number of users, and \(  n_B^i \) is the $i$-th user's batch number.

To achieve effective unlearning, we minimize the error between the noise matrix and the forgotten class, ensuring that the noise matrix adequately replaces the deleted data \( D_f \). This approach enables zero-shot federated unlearning, where users send the trained noise matrix $N_f$ to the server in place of the actual data, preserving data privacy while facilitating the unlearning process. 

In multi-class data scenarios, users can independently train a noise matrix for each class of data they wish to forget.
These noise matrices are then transmitted to the server, where they are integrated into the unlearning process. This approach ensures both privacy preservation and compliance with unlearning requests across multiple data classes. 

\subsection{Knowledge Disentanglement of Forgotten Data}

During the unlearning process,  removing the forgetting data \( D_f \) 
can inadvertently lead to the knowledge loss of the remaining dataset $D_r$ if there is knowledge entanglement between the two. 
To mitigate this issue, it is essential to disentangle \( D_f \) from \( D_r \), ensuring that the integrity of \( D_r \) is preserved.

Liang \textit{et al.} \cite{liang2020training} and Lin \textit{et al.} \cite{lin2023erm} introduced the use of a mask vector applied after the last convolutional layer to achieve class-specific feature representations.  
Their optimization objective promotes sparsity in the mask vectors for each class. Inspired by this approach, we regularize the output \( F^{conv} \in \mathbb{R}^{C \times H \times W} \) of the last convolutional layer for the input \( D_f \) to the model. 
To determine the importance of each channel, we calculate the L1 norm of its output feature map \( F_i^{conv} \in \mathbb{R}^{H \times W} \), where \( i \in C \). Channels with larger L1 norms are considered more representative of the data from  \( D_f \). The disentangling process involves suppressing the outputs of less important channels based on their L1 norms.
Using a specified channel retention ratio \( \alpha \in (0,1) \), we compute a threshold \( Thr \) to retain only the top \( \alpha \) proportion of channels out of the total \( C \) channels. Channels with L1 norm values below  \( Thr \) are suppressed. This process ensures the effective separation of \( D_f \)'s knowledge from the rest of the data. The procedure is mathematically described by Equations \ref{threshold} and \ref{disentangle loss}:

\begin{equation}
    Thr = \text{Threshold}(\alpha, F^{conv})
    \label{threshold}
\end{equation}
\begin{equation}
    \mathcal{L}_{\text{disentangle}} = \frac{1}{(1-\alpha)C} \sum_{i=1}^{C} \text{norms}(F_i^{conv})
    \label{disentangle loss}
\end{equation}

where 

\begin{equation}
    \text{norms}(F_i^{conv}) = \begin{cases} 
\|F_i^{conv}\|_1 & \text{if } \|F_i^{conv}\|_1 < Thr \\
0 & \text{otherwise}
\end{cases}
\end{equation}

By minimizing the loss \( \mathcal{L}_{\text{disentangle}} \), the model undergoes knowledge disentanglement, effectively reducing the entanglement of knowledge related to \( D_f \). This disentanglement process prepares the model for a more targeted and efficient forgetting process, ensuring that the influence of the forgotten data is minimized while preserving the integrity of the remaining knowledge. 

\subsection{Loss Function Construction}
The conventional approximate unlearning approach operates by building the global model from the previous iteration and selectively eliminating knowledge related to the deleted data in response to user deletion requests. 
The unlearning goal is to ensure that, after class-specific forgetting, the model behaves as if it had never been exposed to the deleted class data when processing future inputs. 

However, following the recent recommendation of  \cite{cotogni2023duck, huang2021evaluating, sun2023generative},   after forgetting, a model should not revert to making random predictions. Instead, it should classify the forgotten samples into the most semantically similar remaining categories.
Specifically, for each sample $(x_f, y_f) \in D_f$, we first introduce a hard loss $\mathcal{L}_{\text{hard}}$ 
that penalizes the correct classification of the forgotten category. 
This encourages the model to revise its predictions and shift away from the forgotten class 
\cite{graves2021amnesiac, jang2022knowledge}.
To further guide the unlearning process, we introduce a confusion loss  $\mathcal{L}_{\text{confusion}}$, 
which leverages inter-class similarity to guide the model toward 
plausible alternative categories.   

Additionally, we incorporate a distillation loss  $\mathcal{L}_{\text{distillation}}$, inspired by the method in \cite{chundawat2023can},  to introduce a degree of randomness.
In this setup,  an incompetent teacher model $M_{{bad}}$, which lacks exposure to the forgotten data,  guides the unlearning of a student model $M$. Knowledge distillation is performed such that the student model's output on $D_f$ aligns with that of the teacher. This approach ensures that only residual, non-target knowledge is retained by the student after unlearning. 

To precisely define the concept of guidance, we formulate our initial loss function as shown in  Equation \ref{total unlearn loss}:
\begin{equation}
    \mathcal{L}_{\text{unlearn}} = \mathcal{L}_{\text{hard}} + \mu_c \mathcal{L}_{\text{confusion}}+ \mu_d \mathcal{L}_{\text{distillation}} 
    \label{total unlearn loss}
\end{equation}
where the unlearning loss \( \mathcal{L}_{\text{unlearn}} \) is defined as the weighted sum of three components: 
the hard loss \( \mathcal{L}_{\text{hard}} \), which measures the discrepancy between the output of the student model and the ground truth labels of the removed data;
the confusion loss \( \mathcal{L}_\text{confusion} \), which captures the divergence between the student model's output and the most semantically similar but incorrect class labels; 
and the distillation loss \( \mathcal{L}_\text{distillation} \), which enforces consistency between the outputs of the teacher model and the student model on the removed data.
Constants $\mu_c$ and $\mu_d$ are hyperparameters that control the trade-off between confusion and distillation relative to the primary forgetting objective. 
{``Forgetting" objective is driven by \( \mathcal{L}_{\text{hard}} \) , which is therefore assigned the highest base weight. \( \mathcal{L}_\text{confusion} \) and \( \mathcal{L}_\text{distillation} \) act as auxiliary regularizers to guide the forgetting process more effectively, warranting lower weights.}
We elaborate on each of these loss components in the following paragraphs.

\textit{Hard Loss. }
Hard Loss quantifies the extent to which a model is penalized for correctly predicting data that has been designated for forgetting. 
During standard training, the model learns to minimize the loss across all samples, thus acquiring knowledge from the data. In contrast, the goal of unlearning is to reverse this process, that is, to maximize the model's loss on the forgotten samples, effectively reducing its confidence and accuracy on them.
To achieve this, we employ a cross-entropy loss function, as shown 
in Equation \ref{eq:hard_loss}, where \(M(x_f)\) denotes the
confidence with which the student model predicts the input feature vector \(x_f\) as \(y_f\). Unlike the training process, \(\mathcal{L}_\text{hard}\) aims to reduce the accuracy of the student model on the forgotten samples.

\begin{equation}
\label{eq:hard_loss}
    \mathcal{L}_\text{hard} = \sum_{(x_f, y_f) \in D_f} y_f \log M(x_f)
\end{equation}

\textit{ Confusion Loss. }
The Confusion Loss $\mathcal{L}_{\text{confusion}}$ is based on the model's predicted probability distribution to find the closest non-correct class label for each sample. 
Specifically, for each sample $(x_f, y_f) \in D_f$, we define the confidence of the global model $M$ for the sample $x_f$ as $P_{x_f}^{M} \in \mathbb{R}^{N_{\text{class}}}$, where $N_{\text{class}}$ is the class number and $P_{x_f}^{M}$ can be calculated using the Equation \ref{C_o output}:

\begin{equation}
\label{C_o output}
    P_{x_f}^{M} = \frac{\exp\left({z_i}\right)}{\sum_{j=1}^{N_{\text{class}}} \exp\left({z_j}\right)}
\end{equation}

where $z_i$ represents the confidence of the model $M$ in predicting $x_f$ as correct label $i$ (where $i \in [1, N_{\text{class}}]$), and $z_j$ represents the confidence of the model $M$ in predicting $x_f$ as label $j$ (where $j \in [1, N_{\text{class}}]$). We find the closest non-correct class label $y_{{fake}}$ to $x_f$ based on the predicted probabilities of each class in $P_{x_f}^{M}$, as shown in Equation \ref{closest non-correct class label}:
\begin{equation}
    \label{closest non-correct class label}
    y_{\text{fake}} = \arg\max_{i \neq y_f} P_{x_f}^{M}(i) \quad \text{where} \quad i \in [1, N_{\text{class}}]
\end{equation}

Finally, we use the cross-entropy loss function to align $x_f$'s output $M(x_f)$ with its corresponding $y_{\text{fake}}$,  causing the model's decision boundary to change and merging $x_f$ into the class $y_{\text{fake}}$. The loss function is expressed as Equation \ref{confusion loss}:

\begin{equation}
\label{confusion loss}
    \mathcal{L}_{\text{confusion}} =-\sum_{(x_f, y_f) \in D_f} y_\text{fake} \log M(x_f)
\end{equation}

\textit{Distillation Loss. }
Considering that some samples may encounter difficulties when finding $y_{\text{fake}}$, where the confidence in non-correct classes is evenly matched, this situation could lead to multiple possible $y_{\text{fake}}$, resulting in an unclear forgetting target during the actual unlearning process. 
Consequently, this may cause the model to struggle with convergence. Therefore, we introduce a distillation loss to encourage randomness in the model's predictions.

We use the incompetent teacher model $M_{{bad}}$'s output as the label for the student model $M$. The teacher model's output vector is transformed into a prediction confidence vector through the softmax function. The confidence of \( M_{{bad}} \) for sample \( x_f \) is represented as \( P_{x_f}^{M_{{bad}}} \in \mathbb{R}^{N_{\text{class}}}\), which is calculated by Equation \ref{teacher output}:
\begin{equation}
    P_{x_f}^{M_{{bad}}} = \frac{\exp\left(\frac{v_i}{{Temp}}\right)}{\sum_{j=1}^{N_\text{class}} \exp\left(\frac{v_j}{{Temp}}\right)}
    \label{teacher output}
\end{equation}

where \( {Temp} \) represents the distillation temperature, \( v_i \) indicates the teacher model \( M_{{bad}} \) predicting \( x_f \) as label \( i \) (where \( i \in [1, N_{class}] \) ), \( v_j \) indicates the teacher model \( M_{{bad}} \) predicting \( x_f \) as label \( j \) (where \( j \in [1, N_{class}] \) ). Similarly, we define the confidence of the student model \( M \) for sample \( x_f \) as \( P_{x_f}^M \), which can be calculated by  Equation \ref{student output}:

\begin{equation}
    P_{x_f}^M = \frac{\exp\left(\frac{z_i}{{Temp}}\right)}{\sum_{j=1}^{N_\text{class}} \exp\left(\frac{z_j}{{Temp}}\right)}
    \label{student output}
\end{equation}

where \( z_i \) indicates the student model \( M \) predicting \( x_f \) as label \( i \) (where \( i \in [1, N_\text{class}] \)), and \( z_j \) indicates the student model \( M \) predicting \( x_f \) as label \( j \) (where \( j \in [1, N_\text{class}] \) ). Finally, we define \( \mathcal{L}_{\text{distillation}} \) as follows:

\begin{equation}
    \mathcal{L}_{\text{distillation}} = D_{KL}(P_{x_f}^{M_{{bad}}}|| P_{x_f}^M) =\frac{1}{|D_f|} \sum_{x_f \in D_f} P_{x_f}^{M_{{bad}}} \log \frac{P_{x_f}^{M_{{bad}}}}{P_{x_f}^M}
    \label{dis loss 1}
\end{equation}

The formula shows that the greater the difference in prediction distributions between the teacher and student models on the forgotten dataset, the larger the loss value. 

To reduce the bias caused by a single teacher model, we introduce a set of teacher models \( T_{{set}} = \{T_i\}_{i=1}^{N_T} \), where \( N_T \) is the number of teacher models used. Finally, \( \mathcal{L}_{\text{distillation}} \) is expressed as Equation \ref{dis loss 2}:

\begin{equation}
    \mathcal{L}_{\text{distillation}} = \frac{1}{N_T}D_{KL}(P_{x_f}^{M_{{bad}}}|| P_{x_f}^M) =\frac{1}{N_T}\frac{1}{|D_f|} \sum_{x_f \in D_f} P_{x_f}^{M_{{bad}}} \log \frac{P_{x_f}^{M_{{bad}}}}{P_{x_f}^M}
    \label{dis loss 2}
\end{equation}

In addition to the three primary losses mentioned above, we further enrich and optimize the overall loss function by incorporating model weight drift loss, gradient harmonization, and gradient masking.

\textit{Model Weight Drift Loss. }
During the unlearning process, a decline in the model's performance is inevitable. 
Starting with the original model, which has already reached convergence, the ideally unlearned model should remain close to the original model in parameter space to preserve the same prediction performance on the remaining data. Therefore, we define the model parameter drift loss as shown in Equation \ref{drift loss}. 
\begin{equation}
    \mathcal{L}_{\text{drift}} = \frac{1}{2} \|\omega^{{un}} - \omega^{t}\|_2^2
    \label{drift loss}
\end{equation}
where \( \omega^{{un}} \) represents the model parameters during unlearning, and \( \omega^{{t}} \) represents the parameters before unlearning. By minimizing the change in model parameters, we encourage the model parameters to remain as close as possible to the original model in the parameter space during the unlearning process. We achieve the ``forgetting" task of the model by minimizing the \( \mathcal{L}_{\text{unlearn}} \) loss and the ``remembering" task by minimizing the \( \mathcal{L}_{\text{drift}} \). Therefore, the total loss \( \mathcal{L} \) during the model optimization process can be expressed as Equation \ref{total loss}:

\begin{equation}
    \mathcal{L} = \mathcal{L}_{\text{unlearn}} + \mathcal{L}_{\text{drift}}
    \label{total loss}
\end{equation}

The update of model parameters can be expressed as:
\begin{equation}
    \omega^{{un}} = \omega^{un} - \mu \nabla_{\omega} \mathcal{L} 
 = \omega^{un} - \mu \nabla_{\omega} \mathcal{L}_{\text{unlearn}} - \mu \nabla_{\omega} \mathcal{L}_{\text{drift}}
\end{equation}
where \( \mu \) is the learning rate during the model update process. 
For clarity in the following explanation, we denote the gradient of \( \mathcal{L}_{\text{unlearn}} \) with respect to the model parameters as \( g_f \), representing the gradient generated by the ``forgetting" task, and the gradient of \( \mathcal{L}_{\text{drift}} \) with respect to the model parameters as \( g_r \), representing the gradient generated by the ``remembering" task.

\textit{Gradient Harmonization. }
In the process of unlearning, forgetting and remembering typically represent two conflicting optimization paths. Optimizing a model to remember generalized knowledge may inadvertently cause the model to retain all knowledge, including knowledge that should be forgotten. In contrast, prioritizing forgetting might prevent the model from maintaining valuable generalized knowledge. This can lead to conflicts among gradients, thus diminishing the effectiveness of the overall optimization framework.

Therefore, inspired by previous work in multitask learning \cite{pan2023gradmdm, yu2020gradient}, we employ a gradient harmonization strategy that harmonizes the gradient directions of ``remembering" and ``forgetting" through gradient projection\cite{huang2024learning}. This gradient harmonization strategy effectively reduces gradient conflicts, thereby achieving a consistent and efficient optimization path that satisfies the dual objectives of learning and unlearning.

To be more precise, when dealing with two gradient vectors, one resulting from the forgetting task (\(g_f\)) and the other from the remembering task (\(g_r\)), these gradients often clash during the optimization process. A common approach to simultaneously optimizing both tasks is simply adding \(g_r\) and \(g_f\) together to derive the final gradient to update the target model.

However, this method may not be optimal, as conflicting gradients tend to cancel each other out, thereby reducing overall efficiency. To address this issue, we use a gradient projection method to harmonize the gradients associated with our two primary objectives \cite{huang2024learning}, as detailed below.

First, we calculate the cosine similarity between the two gradients, \(\cos(g_r, g_f) = \frac{g_r \cdot g_f}{\|g_r\| \|g_f\|}\). If the cosine similarity is less than zero, it indicates that there are conflicting components between the two gradient vectors. The gradient harmonization strategy involves computing the projection of the gradient vector \(g_f\) in the direction of \(g_r\). The conflicting component can be expressed as \(\frac{g_r \cdot g_f}{\|g_r\|} \frac{g_r}{\|g_r\|}\), and by subtracting this component from \(g_f\), we obtain the projection gradient of \(g_f\) in the direction orthogonal to \(g_r\). If the cosine similarity is greater than or equal to zero, it suggests that \(g_r\) and \(g_f\) are not in conflict, and thus gradient harmonization is unnecessary. The corrected gradient \(g_f'\) can be represented by  Equation \ref{gf'}:
\begin{equation}
    g_f' = \begin{cases} 
g_f - \frac{g_r \cdot g_f}{\|g_r\|} \frac{g_r}{\|g_r\|} & \text{if } \cos(g_r, g_f) < 0 \\
g_f & \text{otherwise}
\end{cases}
\label{gf'}
\end{equation}

To achieve complete ``forgetting" while preserving knowledge that is essential for ``remembering", we employ a method of linearly combining the adjusted gradient vector \( g_f' \) with \( g_r \), resulting in the final composite gradient \( G \), expressed as \( G = g_f' + g_r \). This composite gradient eliminates the conflicting parts between \( g_f \) and \( g_r \). This approach aims to ensure that during the optimization process, we do not disrupt the model's ``memory" of generalized knowledge, but focus on achieving this through ``forgetting" of specific knowledge. In this way, the forgetting feedback becomes more consistent and efficient.

\textit{Gradient Masking}
The gradient mask is designed to prevent the retention of knowledge related to forgotten data \( D_f \) in the model's parameters. Since the model weight drift vector contains information about the original model, which may also include knowledge from the forgotten data, we aim to prevent the gradient \( g_r \) from carrying any information related to \( D_f \). Specifically, we calculate the cross-entropy 
loss \( l(x_f, y_f; \omega^t) = -y_f \log M(x_f; \omega^t) \) using the forgotten data \( (x_f, y_f) \in D_f \) in the original model \( \omega^t \). Taking the deviation of this loss with respect to \( \omega^t \), we obtain the corresponding gradient \( \nabla_{\omega} l(x_f, y_f; \omega^t) \).
To isolate the parameters mostly affected by \( D_f \), we observe that the parameters with larger norm updates in the gradient are more strongly correlated with the forgotten data, which is consistent with  \cite{meerza2024confuse}. We then define a threshold \( \pi \), and for all parameter positions in the model, we apply a gradient mask: we set the mask to 1 for positions where the gradient value is less than the threshold, and 0 for the remaining positions, as shown in Equation \ref{masking}:
\begin{equation}
    m_s = \mathbf{1}(\left| \nabla_{\omega} l((x_f, y_f); \omega^t) \right| < \pi) \quad \text{where} \quad (x_f, y_f) \in D_f
    \label{masking}
\end{equation}
where the notation \(|\cdot|\) represents the operation of taking the absolute value. The resulting mask \(m_s\) has the same size as the model's parameters. To restrict the knowledge from \(D_f\), we perform an element-wise multiplication (also known as the Hadamard product) of \(m_s\) with \(g_r\):

\begin{equation}
    g_r' = g_r \odot m_s
    \label{gradient masking}
\end{equation}

where $\odot$ denotes the element-wise multiplication. By using the masked gradient for model updates, \(g_r'\) will mostly contain the knowledge from \(D_r\). 
\subsection{Repair}

After the unlearning process, the server distributes the unlearned global model to all users. Upon receiving the global model, each client evaluates its performance on their local remaining dataset. Specifically, each user \( i \)  calculates the percentage drop in accuracy on their remaining dataset \( D_r^i \) as shown in Equation \ref{accuracy drop percentage}. 
\begin{equation}
\Delta \text{Acc}^i = \frac{\text{Acc}_t^i - \text{Acc}_{t+1}^i}{\text{Acc}_t^i} \times 100\%
\label{accuracy drop percentage}
\end{equation}
When user \( i \) detects a significant performance drop, the user can initiate the model repair process. The error minimization noise method generates proxy data \( N_r^i \) for its remaining data \( D_r^i \). The user \( i \) trains \( N_r^i \) locally by minimizing the following loss function,  ensuring that the proxy data effectively captures the characteristics of \( D_r^i \) without retaining any information from the deleted data, as shown in Equation \ref{Nrnoise loss}:
\begin{equation}
\label{Nrnoise loss}
\mathcal{L}_{ N_r^i } = \frac{1}{n_B} \sum_{j=1}^{n_B} -y_r^i(j) \log M(N_r^i(j))
\end{equation}
Here, \( y_r^i(j) \) represents the class label of \( D_r^i(j) \), \( n_B \) denotes the batch number of the noise matrix, and \( M(N_r^i(j)) \) indicates the predicted probability distribution obtained by inputting the $j$-th batch of noise into the classifier model targeted for unlearning. 

{ To address the coordination across multiple users in the federated repair process, we detail the following mechanisms. 
(1) Aggregation of $N_r$: When multiple users submit their proxy repair data ($N_r^1, N_r^2, \ldots$), the server aggregates them using a weighted average scheme, similar to the FedAvg algorithm. The weight for each user $i$ is proportional to the size of their remaining dataset $|D_r^i|$, ensuring a balanced contribution that reflects the data distribution. The aggregated proxy data $\overline{N_r}$ is then used for the repair optimization in Equation \ref{Nrnoise loss}. 
(2) Global Repair Scope: The repair process is designed to update the global model. Once the server performs the repair using $\overline{N_r}$, the updated global model $\omega^{t+1}$ is distributed to all clients, not just the ones who triggered the repair. This maintains model consistency across the federation. 
(3) Prevention of Repeated Repair Interference: To prevent unnecessary or conflicting repairs triggered by minor performance fluctuations, a global accuracy drop threshold $\delta$ is established. A repair request from user $i$ is only accepted if $\Delta Acc^i > \delta$. Furthermore, as outlined in Algorithm \ref{alg:ZeroShotUnlearning} (lines 21-28), the repair is an optional, one-time procedure following the unlearning step, which effectively avoids repeated repair cycles and potential interference between different users' $N_r$.}
 
In addition to the aspects mentioned above, it is crucial to recognize that during the federated learning process, malicious clients may submit harmful model updates, potentially causing significant damage to model performance \cite{yin2018byzantine,li2022improved}.
Various defense mechanisms can be employed to detect and mitigate the impact of these malicious updates. These strategies typically involve inspecting or directly computing with model weights, such as utilizing output predictions \cite{cao2021provably}, intermediate states (\textit{e.g.}, logits) \cite{rieger2022deepsight}, or different norms (\textit{e.g.}, L2 norm or cosine distance) to assess discrepancies between local models and the global model \cite{nguyen2022flame,fung2020limitations}
Moreover, FreqFed \cite{fereidooni2023freqfed} offers an alternative defense by transforming model updates into the frequency domain to analyze their frequency components. Combined with automated clustering, it can effectively identify and remove potentially malicious updates.
\begin{figure*}[ht]
\centering

    \subfloat[]{\label{testsetmetrics-JSD}\includegraphics[width=0.33\textwidth]{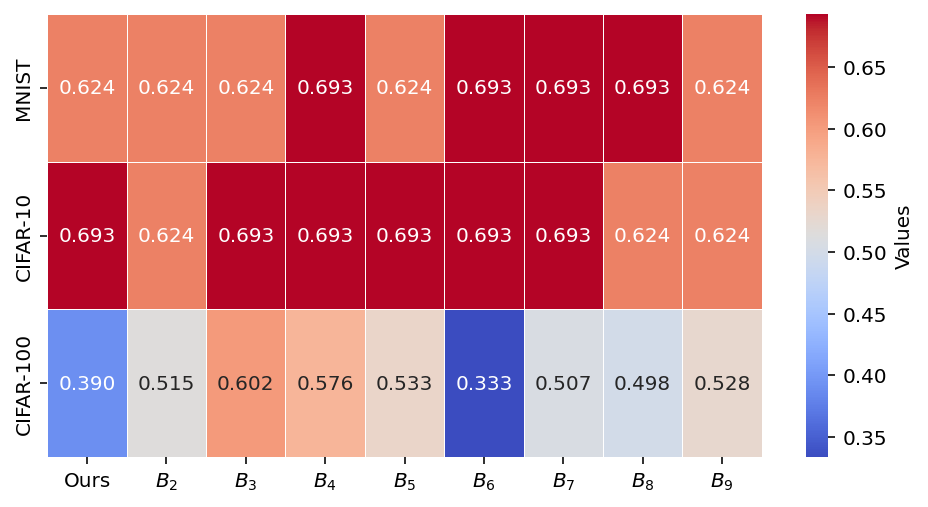}}
    \subfloat[]{\label{testsetmetrics-L2}\includegraphics[width=0.33\textwidth]{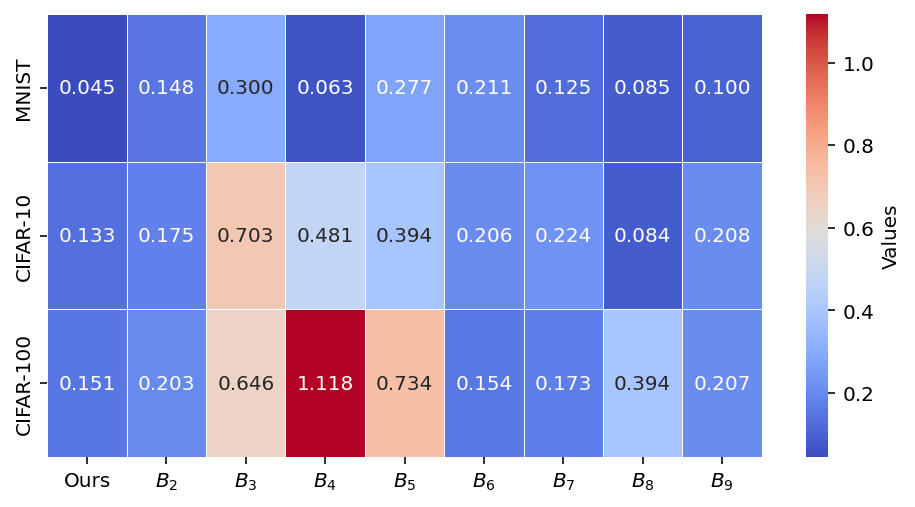}}
    \subfloat[]{\label{testsetmetrics-pvalue}\includegraphics[width=0.33\textwidth]{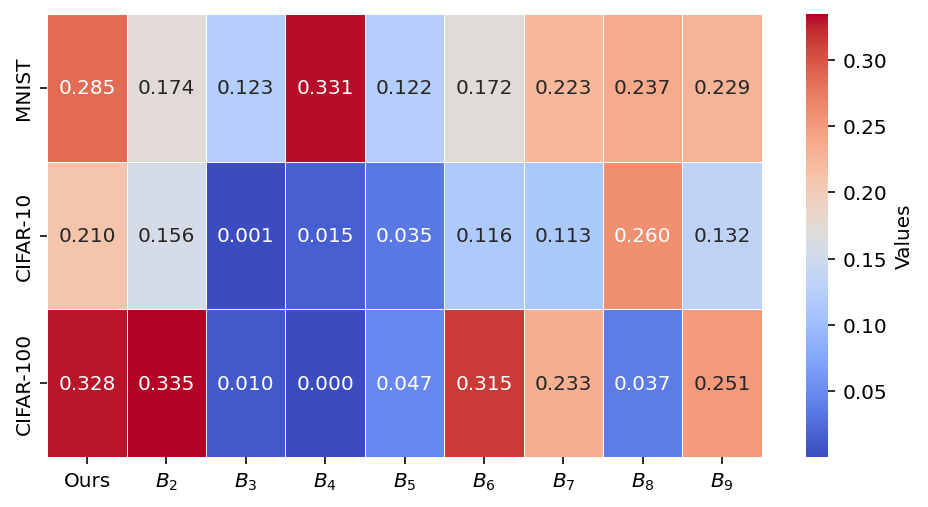}}
 \caption{Performance Metrics on Test Set. (a) JSD on Test Set, (b) L2 on Test Set, (c) p-value on Test Set.}
 \label{fig:JSD,l2,and p}
\end{figure*} 

\section{Experiment}

In this section, 
we begin by outlining the objectives of the experiments.
Next, we describe the experimental setup. 
Finally, we present and analyze the results of the various experiments, offering insights into the performance of our approach compared to the existing methods.

\subsection{Experimental Goal}
Our experimental objective is to validate the effectiveness of the proposed scheme in accomplishing the task of forgetting specific data categories. Subsequently, we aim to assess the contribution of each module within the proposed method.

\begin{enumerate}
    \item Zero-shot federated unlearning. We will conduct comparative experiments to assess whether synthetic data can effectively substitute for actual data in achieving data category forgetting. Specifically, we will examine the difference between the experimental results obtained with real data versus those with proxy data.
    \item Knowledge disentanglement method. We will perform comparative experiments to verify whether the disentangling operation can effectively address the knowledge entanglement between different categories. We will investigate whether the forgetting of categories, after disentangling, leads to a significant collateral decline in the accuracy of other categories, in addition to the target category.
    \item Impact of loss function components. We will explore the impact of different parts of the loss function, including the effects of the proposed loss function components and the gradient harmony mechanism, on the performance of the unlearning process.
    \item Comparison with the state of the art. We will evaluate whether the proposed unlearning method outperforms other existing unlearning techniques.
\end{enumerate}

\subsection{Experimental Setup}
In the experiment, all models are implemented using PyTorch and executed on two machines: one equipped with an NVIDIA 2080 Ti GPU and another with an NVIDIA 4090 D GPU.

\textbf{Dataset Description}.
In the experiments, we utilized three publicly available ML datasets: MNIST \cite{lecun1998gradient}, CIFAR-10 \cite{krizhevsky2009learning}, and CIFAR-100 \cite{krizhevsky2009learning}. As shown in Table \ref{tab:dataset}, these datasets encompass varying attributes, dimensions, and the number of categories.

\begin{table}[h]
    \centering
    \renewcommand\arraystretch{1.5}
    \caption{Dataset Description}
    \label{tab:dataset}
    \begin{tabular}{ccccc}
    \hline
    Dataset       & Dimensions & Classes & Training & Test  \\ \hline
    MNIST         & 784        & 10      & 60000    & 10000 \\
    
    CIFAR-10      & 3072       & 10      & 50000    & 10000 \\ 
    {CIFAR-100}     & {3072}       & {100}     & {50000}    & {10000} \\ \hline
    \end{tabular}
\end{table}

\textbf{Models}.
Following the most related works \cite{chundawat2023can,hitaj2017deep, gong2023redeem}, we adopt four different models for our evaluation. 
In particular, the model for MNIST is a traditional LeNet-5 model \cite{liu2022right} \cite{lecun1998gradient}, which consists of 2 convolution layers, 2 max pool layers, and 2 fully connected layers for prediction output.
The model selected for CIFAR-10 is ResNet32, a variant of the Residual Network (ResNet) architecture proposed by He \textit{et al.} \cite{he2016deep}. It comprises 32 layers, structured with multiple residual blocks that enable the network to learn residual functions relative to its input.
The model chosen for CIFAR-100 is ResNet56, an advanced variant of ResNet architecture introduced by He \textit{et al.} \cite{he2016deep}. This model, tailored for the CIFAR-100 dataset, comprises 56 layers with multiple residual blocks.

\textbf{Hyperparameters}.
Our hyperparameter settings are aligned with those of Hitaj \textit{et al.} \cite{hitaj2017deep}.
For the MNIST dataset, we utilize a batch size of 100, a learning rate of 0.01, and the SGD optimizer.
For the CIFAR-10 and CIFAR-100 datasets, we use a batch size of 128, a learning rate of 0.01, and the Adam optimizer.
In the category unlearning task, following \cite{chundawat2023can}, we set the number of categories to be forgotten to either 1 or 20\% of the total category count.
$\alpha$ is set to 0.9.
{ The values of $\mu_c$ and $\mu_d$ were determined through a grid search on a validation set derived from the training data, using a representative scenario involving single-class unlearning on CIFAR-10. Both parameters are ultimately set to 0.5.}

\textbf{Evaluation Metrics}.
Based on the experimental setup of \cite{chundawat2023can, cotogni2023duck, huang2025learning}, we measure the effectiveness of different forgetting learning methods using precision on the remaining dataset and precision on the forgotten dataset.

Furthermore, we use L2 distance, Jensen-Shannon Divergence (JSD), and T-test to quantify the similarity between the model after applying the proposed unlearning approach and the retrained model. A smaller JSD and L2 distance indicates a higher similarity between the two distributions.
A T-test is a statistical method used to determine if there is a significant difference between the means of two groups. In a T-test, the p-value (probability value) indicates the probability of obtaining the observed results (or more extreme) if the null hypothesis (\textit{i.e.}, the means of two groups are equal) is true. The smaller the p-value, the stronger the evidence against the null hypothesis, indicating a greater likelihood of a significant difference between the means of the two samples.

\textbf{Baselines}. 
To compare with the most recent and relevant work and demonstrate the effectiveness of the proposed approach, we establish the following baselines.
The first baseline, denoted as $B_1$, retrains the model from scratch \cite{zhang2023fedrecovery}. It is the golden baseline of machine unlearning.
The second baseline, denoted as $B_2$, is the original model that is fine-tuned on the forget-set $D_f$ following the negative direction of the gradient descent\cite{golatkar2020eternal}.
The third baseline, denoted as $B_3$, is the original model that is fine-tuned with the forget-set $D_f$, randomly selecting a label to compute the cross-entropy loss function\cite{golatkar2020eternal}.
The fourth baseline, denoted as $B_4$, is the original model that is fine-tuned only on the retain-set with a large learning rate to remove the knowledge on the forget-set and maximize the accuracy on the retain-set\cite{golatkar2020eternal}.
$B_2$, $B_3$, and $B_4$ are the most classic and commonly used baseline methods in the field of machine unlearning.
They can also be adapted for use in federated unlearning scenarios.
The fifth baseline, denoted as $B_5$, employs an incompetent teacher model similar to our unlearning approach \cite{chundawat2023can}.
The sixth baseline, denoted as $B_6$, CONFUSE\cite{meerza2024confuse}, designs a Confusion Loss and performs Saliency-guided federated unlearning.
The seventh baseline, denoted as $B_7$, QuickDrop\cite{dhasade2024quickdrop}, reverse training (stochastic GA) on the distilled dataset, and fine-tuning with a few original data samples. It considers the zero-sample unlearning process in federated unlearning.
$B_6$ and $B_7$ are the most advanced methods in the field of federated unlearning.
The eighth baseline, denoted as $B_8$, SCalable Remembering and Unlearning unBound (SCRUB)\cite{kurmanji2024towards}, is the state-of-the-art unlearning method for deep learning settings, which uses a teacher-student network to address the forgetting problem. 
The ninth baseline, denoted as $B_9$, NegGrad+\cite{kurmanji2024towards}, is a fine-tuning-based unlearning approach that achieves precise forgetting through a multitask setting of the loss function.
\begin{figure}[t]
\centering
\subfloat[]{\label{remaining data TP}\includegraphics[width=0.45\textwidth]{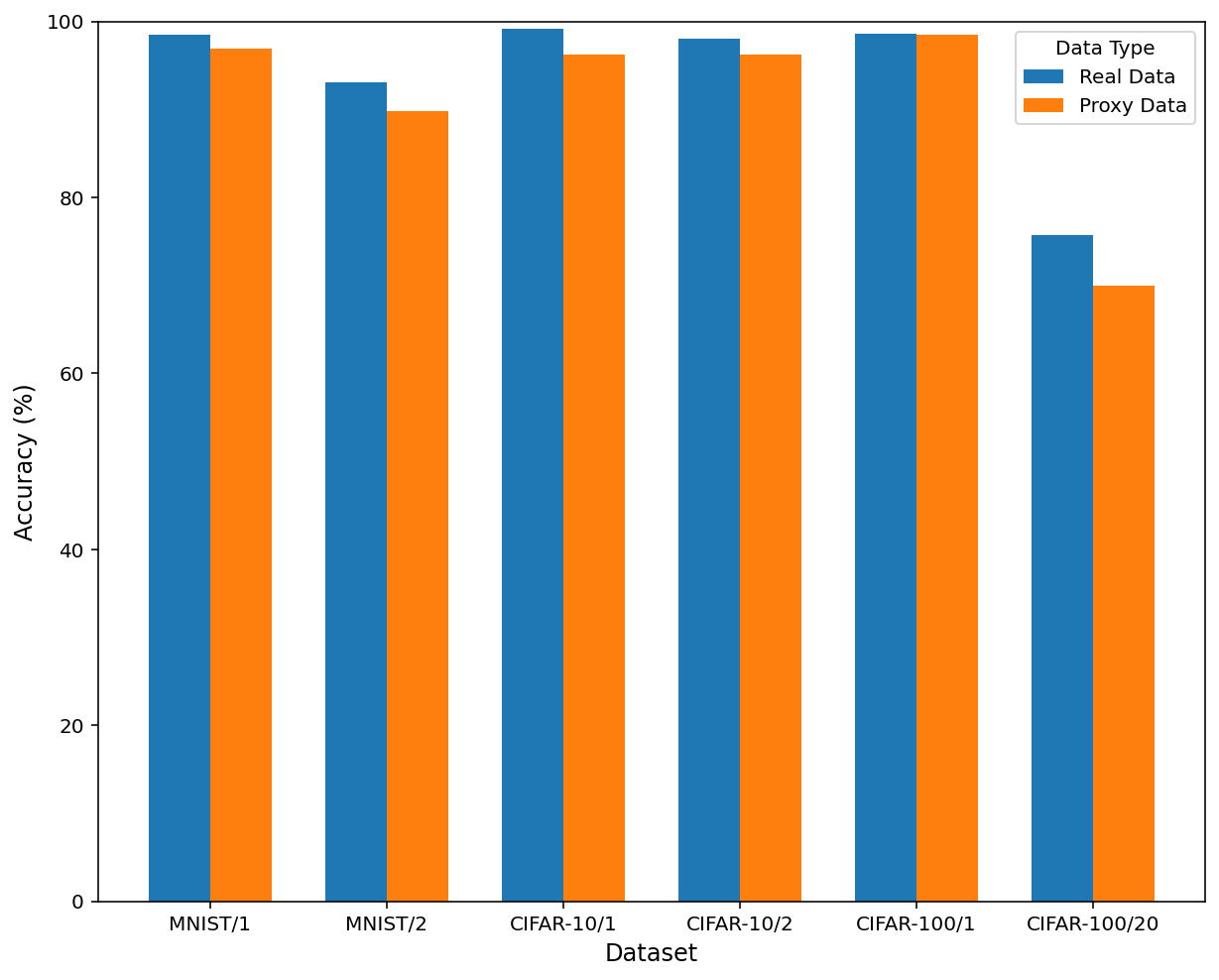}}
    \subfloat[]{\label{forgetting data TP}\includegraphics[width=0.45\textwidth]{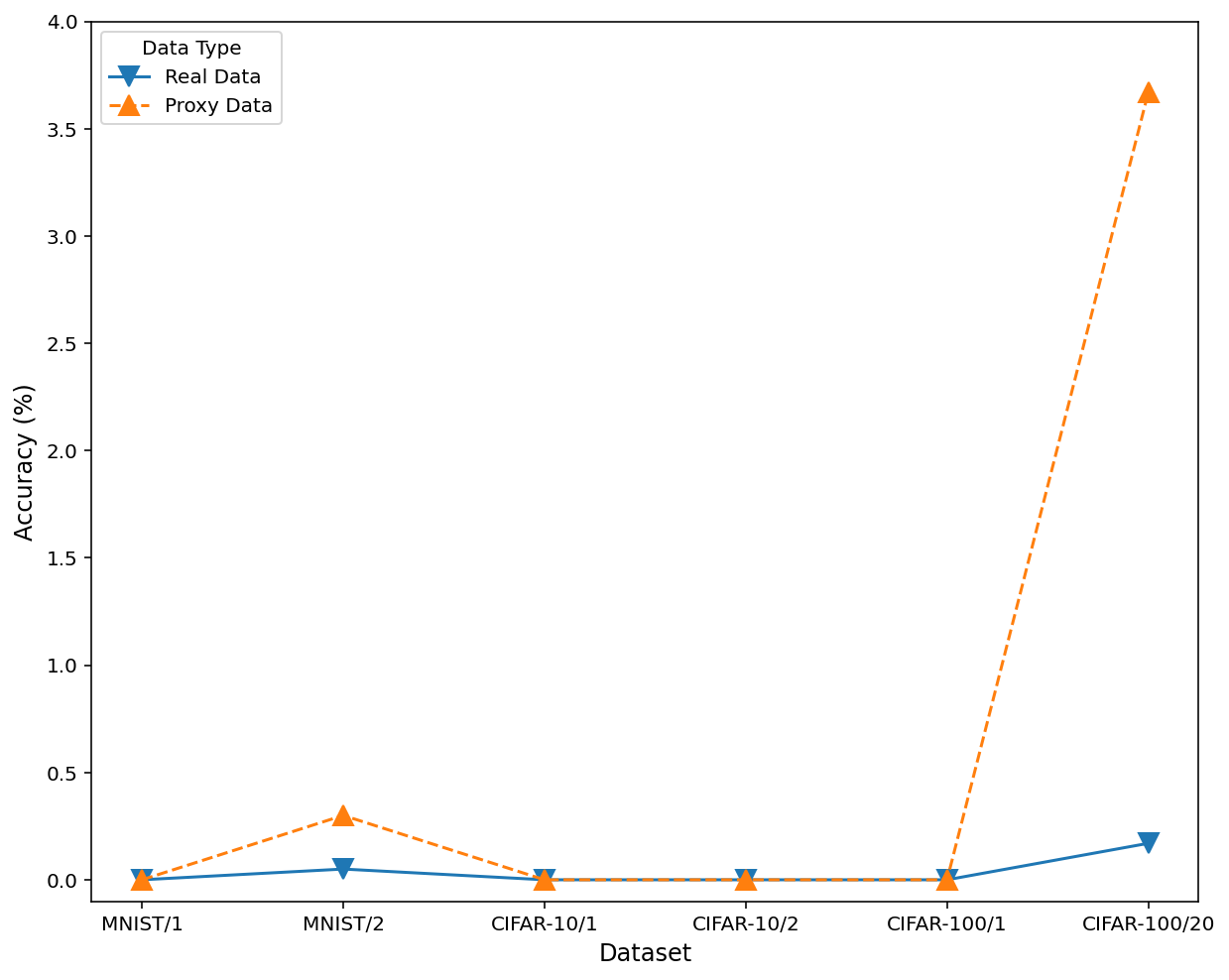}}
\caption{Comparison of accuracy on remaining and forgotten data using real and proxy data for unlearning. (a) accuracy on remaining data, and (b) accuracy on forgotten data.}
\label{fig: true and fake}
\end{figure}
\begin{figure}[t]
\centering
\includegraphics[width=0.9\linewidth]{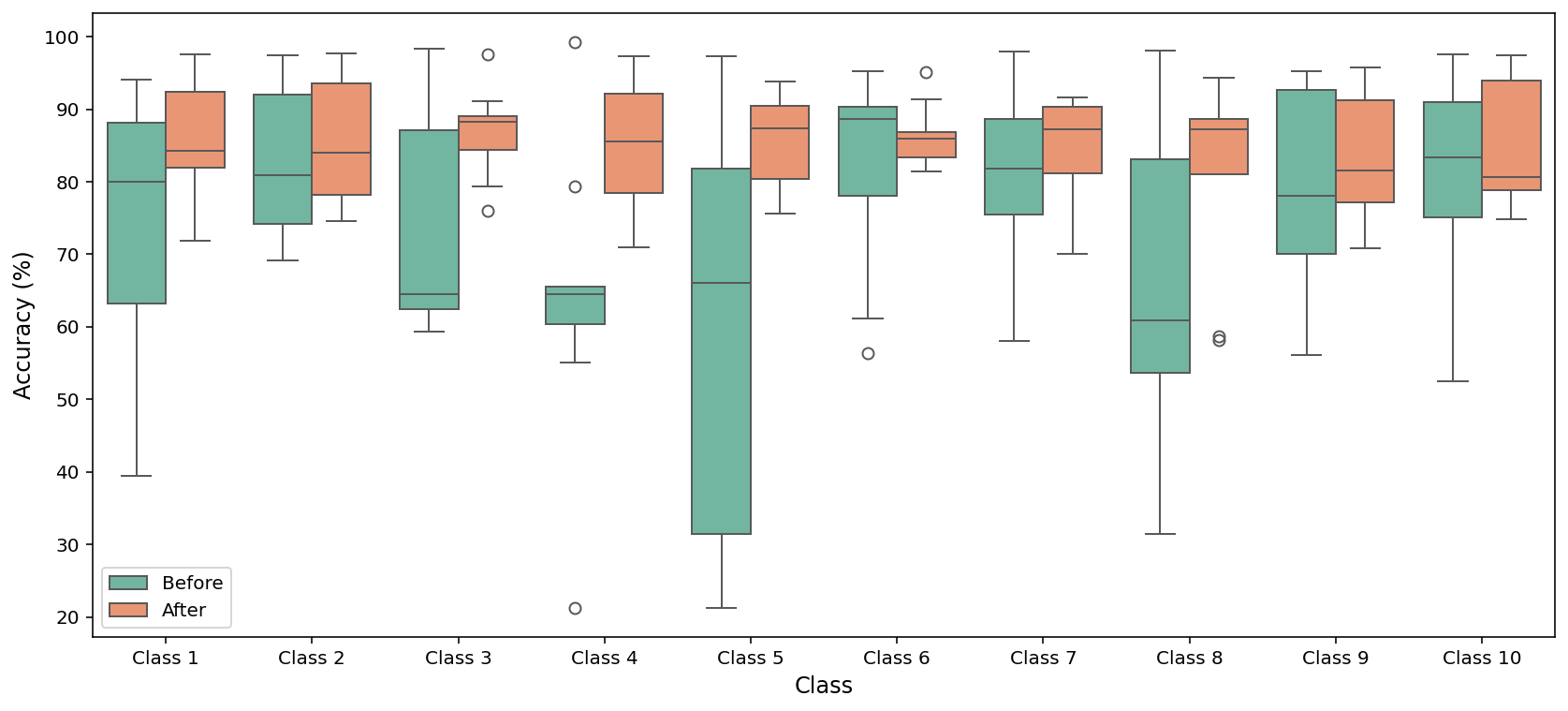}
\caption{Accuracy of non-target categories before and after disentangling.}
\label{fig:decoupling}
\end{figure}

{
\subsubsection{Baseline Implementation Details}

The baseline methods $B_2$, $B_3$, $B_4$, $B_7$,$ B_8$, and $B_9$ were originally designed for centralized settings, requiring direct access to the forget set $D_f$ or its gradients on the server. To enable comparison in the federated learning context, we adapted these methods by migrating their unlearning processes to the server-only. 

For methods $B_5$ and $B_6$, which were specifically designed for federated learning, we implemented them strictly according to their original papers. Their unlearning processes are executed on the client-side, with only model updates being aggregated on the server, thus fully complying with the zero-shot threat model requirements.

The retraining method ($B_1$) serves as the performance upper bound, where the model is trained on the server using the complete original data $D_f$. This configuration explicitly violates the zero-shot threat model, but its results provide a crucial theoretical performance ceiling for evaluating the effectiveness of other methods.

This experimental design aims to establish a comprehensive and rigorous performance benchmark. Through the transparent adaptation strategies described above, we enable advanced methods originally designed for centralized settings to participate in federated environment evaluations. The experimental results demonstrate that the Jellyfish method, while strictly adhering to the zero-shot threat model (where the server cannot access any form of raw or proxy data), remains highly competitive compared to these advanced baseline methods operating under relaxed assumptions. This strongly validates the practical value and advantages of Jellyfish in privacy-preserving federated unlearning.}

\subsection{Experimental Result}

\textbf{Unlearning Approach Evaluation}.
For the unlearning task, we follow the methodology outlined in \cite{chundawat2023can}, using the MNIST, CIFAR-10, and CIFAR-100 datasets to study category unlearning. Specifically, we focus on single-category unlearning and randomly select 20\% of the total categories for unlearning\cite{chundawat2023can}. The number of categories in the specific forgetting process is denoted as $Num_{f}$.
We measure the model's test accuracy separately for the remaining and unlearned categories.

{ 
To verify whether the Jellyfish scheme genuinely eliminates generalizable knowledge about the forgotten data $D_f$ rather than merely obscuring the memorization of training instances, we focus on analyzing the model's behavior on the test set. Performance metrics on the training set will be provided as supplementary material and moved to the appendix \ref{apd:A} for reference.
}

{

\begin{sidewaystable}
\renewcommand\arraystretch{2}
\tabcolsep=0.1cm
\caption{Category unlearning on MNIST, CIFAR-10, and CIFAR-100 datasets.}
\centering
\begin{tabular}{cccccccccccccc}
\hline
Dataset                    & $Num_{f}$           & Metrics      & origin   & $B_1$   & $B_2$   & $B_3$   & $B_4$   & $B_5$   & $B_6$   & $B_7$   & $B_8$   & $B_9$   & Ours    \\ \hline
\multirow{4}{*}{MNIST}     & \multirow{2}{*}{1}  & acc on $D_r$ & 98.90\%  & 99.04\% & 93.11\% & 92.71\% & 95.26\% & 93.13\% & 95.08\% & 96.50\% & 96.76\% & 96.54\% & 97.85\% \\
                           &                     & acc on $D_f$ & 99.59\%  & 0.00\%  & 0.00\%  & 0.00\%  & 6.26\%  & 0.00\%  & 1.73\%  & 0.31\%  & 0.00\%  & 0.00\%  & 0.00\%  \\
                           & \multirow{2}{*}{2}  & acc on $D_r$ & 99.06\%  & 97.53\% & 92.79\% & 85.84\% & 87.82\% & 89.71\% & 89.34\% & 90.34\% & 91.28\% & 87.83\% & 90.84\% \\
                           &                     & acc on $D_f$ & 99.38\%  & 0.00\%  & 0.16\%  & 1.04\%  & 4.68\%  & 2.46\%  & 4.11\%  & 0.24\%  & 0.03\%  & 0.00\%  & 0.00\%  \\ 
\cline{1-14}
\multirow{4}{*}{CIFAR-10}  & \multirow{2}{*}{1}  & acc on $D_r$ & 90.13\%  & 91.70\% & 85.73\% & 79.78\% & 68.11\% & 87.56\% & 85.71\% & 84.10\% & 88.50\% & 84.78\% & 87.23\% \\
                           &                     & acc on $D_f$ & 94.60\%  & 0.00\%  & 0.60\%  & 4.94\%  & 2.50\%  & 7.57\%  & 9.00\%  & 0.20\%  & 1.10\%  & 0.00\%  & 0.00\%  \\
                           & \multirow{2}{*}{2}  & acc on $D_r$ & 90.13\%  & 90.10\% & 82.06\% & 78.36\% & 86.55\% & 87.10\% & 86.85\% & 85.79\% & 85.99\% & 84.35\% & 87.39\% \\
                           &                     & acc on $D_f$ & 93.55\%  & 0.00\%  & 0.25\%  & 6.95\%  & 7.47\%  & 5.02\%  & 0.65\%  & 1.75\%  & 0.85\%  & 0.00\%  & 0.15\%  \\ 
\cline{1-14}
\multirow{4}{*}{CIFAR-100} & \multirow{2}{*}{1}  & acc on $D_r$ & 64.43\%  & 80.71\% & 63.70\% & 53.58\% & 50.72\% & 62.07\% & 78.24\% & 65.24\% & 64.36\% & 62.91\% & 78.45\% \\
                           &                     & acc on $D_f$ & 84.00\%  & 0.00\%  & 1.00\%  & 5.50\%  & 6.00\%  & 16.00\% & 1.00\%  & 0.70\%  & 0.96\%  & 1.00\%  & 1.00\%  \\
                           & \multirow{2}{*}{20} & acc on $D_r$ & 64.43\%  & 78.13\% & 47.35\% & 47.22\% & 37.67\% & 55.01\% & 58.75\% & 51.96\% & 50.46\% & 50.59\% & 58.79\% \\
                           &                     & acc on $D_f$ & 63.60\%  & 0.00\%  & 0.00\%  & 2.70\%  & 8.60\%  & 19.15\% & 2.90\%  & 1.02\%  & 2.11\%  & 0.61\%  & 1.25\%  \\ \hline
\end{tabular}
\label{acc on Df and Dr test}
\end{sidewaystable}
}

We use the above evaluation metrics to assess the unlearning performance of both the baselines and our method. The experimental results are summarized in  Table \ref{acc on Df and Dr test}.
From Table \ref{acc on Df and Dr test}, our method demonstrates superior performance compared to other baselines in deleted data. Moreover, our approach achieves performance comparable to $B_1$ in the remaining data.

Additionally, we employed Jensen-Shannon Divergence (JSD), L2 distance, and T-test to quantify the similarity between the model after applying the proposed unlearning approach and the retrained model on test set. The results are shown in Figure \ref{fig:JSD,l2,and p}. 
From the figure,  we can learn that our approach has smaller L2 values. Moreover, our method has a lower JSD value compared to other baselines, indicating that the predictive results obtained through our method are closer to those obtained through $B_1$. 
In the context of the T-test, our algorithm consistently yields smaller p-values in most cases. This suggests significant differences between the predictive patterns obtained through our algorithm and those generated by the original model.

{
Furthermore, to substantiate the privacy guarantees of the proposed unlearning scheme, we evaluated its resilience against Membership Inference Attacks (MIA), a standard method for quantifying privacy risk\cite{hu2022membership}. The MIA success rates on both $D_r$ and $D_f$ are summarized in Table \ref{tab:mia}.
The experimental results demonstrate that Jellyfish effectively mitigates privacy risks. On $D_f$, the MIA success rate drops dramatically to a level close to 50\%---equivalent to random guessing---and aligns closely with the retrained model ($B_1$). This indicates that the unlearned model successfully removes the membership information of $D_f$, making it statistically indistinguishable from unseen data. Concurrently, the MIA success rate on $D_r$ remains stable, confirming that the unlearning process precisely targets the forgotten data without compromising the privacy or utility of the retained data. These findings provide strong empirical evidence that Jellyfish fulfills the core privacy objective of the ``right to be forgotten" by concretely reducing susceptibility to real-world privacy attacks.
}

\begin{table}[]
\renewcommand\arraystretch{1.5}
\caption{Membership Inference Attack Success Rate.}
\label{tab:mia}
\begin{tabular}{ccccc}
\hline
Dataset                    & Data Partition & Before & Unlearned & Retrained \\ \hline
\multirow{2}{*}{MNIST}     & $D_r$           & 81.13\%  & 79.40\%    & 78.83\%    \\
                           & $D_f$           & 80.57\%  & 50.68\%    & 50.10\%     \\ \hline
\multirow{2}{*}{CIFAR-10}  & $D_r$           & 96.60\%  & 83.45\%    & 86.07\%    \\
                           & $D_f$           & 97.96\%  & 53.28\%    & 50.07\%    \\ \hline
\multirow{2}{*}{CIFAR-100} & $D_r$           & 89.28\%  & 89.61\%    & 89.22\%    \\
                           & $D_f$           & 89.34\%  & 52.55\%    & 51.87\%    \\ \hline
\end{tabular}
\end{table}

{
\textbf{Computational Efficiency Analysis.}
To assess the practical deployability of the Jellyfish scheme in resource-constrained federated learning, we performed a comparative analysis of its computational overhead against two baseline methods, $B_1$ and $B_7$. The evaluation focuses on three core metrics: total execution time, average batch processing time, and GPU memory consumption, as summarized in Table \ref{tab:cost}.

In terms of time efficiency, Jellyfish performs comparably to $B_7$ and significantly outperforms $B_1$ in both total and per-iteration execution time. Regarding GPU memory usage, Jellyfish consumes 4.45 GB on CIFAR-10 (higher than $B_7$'s 3.12 GB) and 1.89 GB on CIFAR-100 (more efficient than $B_7$'s 1.62 GB).

The results confirm Jellyfish's comprehensive advantage in computational efficiency. While $B_7$ excels in individual metrics on certain datasets, Jellyfish demonstrates superior overall execution time. This efficiency is attributed to its knowledge disentanglement and gradient harmonization mechanisms, which enable the target performance to be reached in fewer communication rounds. Notably, on the more complex CIFAR-100 dataset, Jellyfish outperforms the baselines in both total time and memory usage, highlighting its scalability for handling complex models and data distributions.

Analysis reveals that Jellyfish's additional overhead primarily originates from client-side proxy data ($N_f$) generation and server-side knowledge disentanglement. However, this overhead is minimal: the one-time $N_f$ generation is lightweight, as evidenced by Jellyfish's lower total time compared to $B_1$, and disentanglement incurs negligible cost, reflected in the low average batch time. The performance gains from reduced knowledge entanglement far outweigh this modest computational cost.}

\begin{table}[]
\caption{Computational Efficiency Comparison of Different Unlearning Methods.}
\label{tab:cost}
\renewcommand\arraystretch{1.2}
\begin{tabular}{ccccc}
\hline
\multirow{2}{*}{Metric}                         & \multirow{2}{*}{Method} & \multicolumn{3}{c}{Dataset}   \\
                                                &                         & MNIST  & CIFAR-10 & CIFAR-100 \\ \hline
\multirow{3}{*}{Average Batch Time   (s/epoch)} & $B_1$                    & 2.64   & 15.36    & 23.75     \\
                                                & $B_7$                    & 6.91   & 4.51     & 5.14      \\
                                                & Ours                    & 3.25   & 4.49     & 5.62      \\ \hline
\multirow{3}{*}{Total Time (s)}                 & $B_1$                    & 903.27 & 2379.3   & 3081.23   \\
                                                & $B_7$                    & 15.44  & 115.39   & 154.05    \\
                                                & Ours                    & 31.77  & 67.21    & 85.33     \\ \hline
\multirow{3}{*}{GPU Memory (GB)}                & $B_1$                    & 0.98   & 0.95     & 0.64      \\
                                                & $B_7$                    & 1.95   & 3.12     & 1.62      \\
                                                & Ours                    & 2.48   & 4.45     & 1.89      \\ \hline
\end{tabular}
\end{table}

\textbf{Zero-Shot Method Evaluation}.
We compared the accuracy obtained on the remaining data and forgotten data in the testing set when unlearning was performed with real data and proxy data, respectively. The results are shown in Figure \ref{fig: true and fake}, the horizontal coordinate labeled ``MNIST/1" indicates the accuracy when forgetting category number 1 in the MNIST dataset, and so on.
From Figure \ref{fig: true and fake}, we can learn that the results obtained using real and proxy data are not significantly different. The greatest difference is observed when forgetting 20 categories of the CIFAR-100 dataset, but the values are still within an acceptable range. Therefore, this demonstrates that the use of proxy data can achieve zero-shot unlearning tasks.


\textbf{Knowledge Disentanglement Method Evaluation.}
To validate the effectiveness of the proposed disentangling method, we conducted our study on the CIFAR-10 dataset, comparing the accuracy of non-target categories when each category was unlearned. 
Ideally, after decoupling between different categories, the accuracy of non-target categories should not decrease significantly, and the accuracy range of non-target categories should remain within a narrow margin. 
We used box plots to summarize the accuracy of other categories when each category was forgotten, and the results are shown in Figure \ref{fig:decoupling}. From the figure, we can learn that the accuracy after disentangling is more concentrated, with fewer outliers and a higher overall level, proving that the disentangling method has reduced the entanglement of knowledge between categories.

{
\textbf{Ablation Study.}
To comprehensively evaluate the contribution of each component in the proposed Jellyfish framework, we conducted an extensive ablation study on the CIFAR-10 dataset under the setting of forgetting one class. We compared the complete Jellyfish method against several variants: without knowledge disentanglement, without gradient harmonization, without gradient masking, without hard loss, without confusion loss, without distillation loss, without drift loss, and a variant using real data instead of proxy data (Jellyfish-real-data). For each variant, we report the accuracy on \(D_r\) and \(D_f\), as well as the L2 distance and JSD compared to the retrained model (approximation quality). The results are summarized in Table \ref{tab:ablation}.

\begin{table}[h]
\centering
\label{tab:ablation}
\renewcommand\arraystretch{1.5}
\caption{Ablation Study of the Importance of Different Components.}
\label{tab:ablation_study}
\begin{tabular}{lcccc}
\hline
\textbf{Variant} & \textbf{Acc on \(D_r\)} & \textbf{Acc on \(D_f\)} & \textbf{L2} & \textbf{JSD} \\ \hline
Without Disentanglement & 84.02\% & 0.00\% & 0.28344 & 0.63013 \\ 
Without Gradient Harmonization & 75.20\% & 0.00\% & 0.59916 & 0.63013 \\ 
Without Gradient Masking & 90.21\% & 81.10\% & 0.94388 & 0.69315 \\ 
Without Hard Loss & 97.87\% & 32.60\% & 0.93813 & 0.69315 \\ 
Without Confusion Loss & 85.60\% & 0.00\% & 0.22123 & 0.63013 \\ 
Without Distillation Loss & 87.13\% & 0.00\% & 0.17709 & 0.63013 \\ 
Without Drift Loss & 84.02\% & 0.00\% & 0.29615 & 0.60635 \\ 
Jellyfish-real-data & 88.56\% & 0.00\% & 0.10596 & 0.60635 \\ 
Complete Jellyfish & 87.85\% & 0.00\% & 0.12762 & 0.60635 \\ \hline
\end{tabular}
\end{table}

The results demonstrate that the complete Jellyfish method achieves a balance between utility preservation (87.85\% accuracy on \(D_r\)) and effective forgetting (0.00\% accuracy on \(D_f\)), with low L2 distance and JSD to the retrained model. 
The variant without gradient masking exhibits a severe failure in forgetting, with \(D_f\) accuracy remaining at 81.10\%, indicating that gradient masking is crucial for preventing the retention of forgotten knowledge. 
The removal of hard loss also significantly impairs forgetting quality (32.60\% accuracy on \(D_f\)), underscoring its necessity for explicitly penalizing correct predictions on forgotten data. 
The removal of gradient harmonization leads to the lowest utility preservation (75.20\% accuracy on \(D_r\)), highlighting its role in balancing the conflicting objectives of forgetting and remembering. 
Knowledge disentanglement and drift loss contribute to stable utility preservation, as their removal results in a noticeable drop in \(D_r\) accuracy. 
The performance of the Jellyfish-real-data variant is comparable to the complete method, validating the effectiveness of the proposed proxy data approach. 
Components such as confusion loss and distillation loss show relatively minor individual impacts on the primary metrics in this setting but contribute to the overall robustness.

By analyzing the data, the ablation study confirms that the key components of Jellyfish—particularly gradient masking, hard loss, and gradient harmonization—are essential for achieving high-performance federated unlearning. The collective integration of these components enables Jellyfish to effectively forget target data while maintaining model utility and closely approximating the retrained model.}

{
\textbf{Repair Evaluation.}
To validate the zero-shot repair mechanism's effectiveness, we experimented on the CIFAR-10 dataset, focusing on a client that experienced a notable accuracy decline.

The client generated proxy data \(N_r\) for its remaining dataset \(D_r\) using the error-minimization noise technique. The model's performance on the client's local test set was evaluated across four stages: before unlearning, after unlearning, after repair, and after full retraining from scratch. The results are summarized in Table \ref{tab:repair_results}.

\begin{table}[htbp]
\centering
\renewcommand\arraystretch{1.5}
\caption{Evaluation of Repair Mechanism.}
\label{tab:repair_results}
\begin{tabular}{lcc}
\hline
\textbf{Stage} & \textbf{Accuracy} & \textbf{Accuracy Gap vs. Retrained Model} \\
\hline
Before Unlearning & 93.90\% & +1.64\% \\
Unlearned & 85.41\% & -6.85\% \\
Repaired & 91.88\% & -0.38\% \\
Retrained & 92.26\% & 0.00\% \\
\hline
\end{tabular}
\end{table}

}

\section{Conclusion}

In this paper, we propose a zero-shot federated unlearning scheme that enhances the privacy of forgotten data by a novel zero-shot learning mechanism. 
 Additionally, we design a new loss function that integrates the proposed knowledge disentanglement technique and harmonizes the conflicting objectives of forgetting and retaining knowledge through a carefully constructed combination of various losses. To preserve model performance without relying on any remaining data, we further introduce a zero-shot repair mechanism. 
Finally, we conduct comprehensive experiments to validate the effectiveness and robustness of the proposed federated unlearning scheme.


\section*{Declarations}

\subsection*{Funding}
This work is supported by the Strategic Priority Research Program of the Chinese Academy of Sciences, Grant No. XDB0690303.

\subsection*{Conflict of interest/Competing interests}
We have no Conflict of interest.

\subsection*{Code availability}
Our source code is available at \url{https://anonymous.4open.science/r/Jellyfish-B4CD}.
\bibliography{sn-bibliography}

{
\begin{appendices}

\section{Experimental Results on the Training Set}\label{apd:A}

This section presents the experimental results of the unlearning approach evaluation on the training set. Table \ref{acc on Df and Dr train} shows the model's test accuracy is reported separately for $D_r$ and $D_f$ on the training set. Figure \ref{fig:JSD,l2,and p train} illustrates the JSD, L2 distance, and T-test performance metrics evaluated on the test set.

\begin{sidewaystable}
\renewcommand\arraystretch{2}
\tabcolsep=0.1cm
\caption{Category unlearning on MNIST, CIFAR-10, and CIFAR-100 datasets (Train).}
\centering
\begin{tabular}{cccccccccccccc}
\hline
Dataset                    & $Num_{f}$           & Metrics      & origin   & $B_1$   & $B_2$   & $B_3$   & $B_4$   & $B_5$   & $B_6$   & $B_7$   & $B_8$   & $B_9$   & Ours    \\ \hline
\multirow{4}{*}{MNIST}     & \multirow{2}{*}{1}  & acc on $D_r$ & 99.88\%  & 99.80\% & 93.68\% & 93.61\% & 96.31\% & 94.01\% & 96.02\% & 96.20\% & 96.80\% & 97.53\% & 98.48\% \\
                           &                     & acc on $D_f$ & 99.89\%  & 0.00\%  & 0.00\%  & 0.20\%  & 6.50\%  & 0.00\%  & 1.50\%  & 0.45\%  & 0.00\%  & 0.00\%  & 0.00\%  \\
                           & \multirow{2}{*}{2}  & acc on $D_r$ & 99.87\%  & 98.30\% & 93.04\% & 86.18\% & 87.93\% & 90.00\% & 89.29\% & 90.14\% & 91.46\% & 91.06\% & 92.26\% \\
                           &                     & acc on $D_f$ & 99.83\%  & 0.00\%  & 0.87\%  & 1.47\%  & 4.62\%  & 2.97\%  & 4.05\%  & 0.12\%  & 0.00\%  & 0.00\%  & 0.00\%  \\ 
\cline{1-14}
\multirow{4}{*}{CIFAR-10}  & \multirow{2}{*}{1}  & acc on $D_r$ & 99.86\%  & 99.80\% & 95.78\% & 90.56\% & 97.72\% & 96.71\% & 95.69\% & 93.62\% & 97.15\% & 94.62\% & 98.49\% \\
                           &                     & acc on $D_f$ & 100.00\% & 0.00\%  & 0.00\%  & 5.35\%  & 1.57\%  & 7.54\%  & 6.74\%  & 0.00\%  & 0.00\%  & 0.00\%  & 0.00\%  \\
                           & \multirow{2}{*}{2}  & acc on $D_r$ & 99.88\%  & 99.61\% & 90.50\% & 91.83\% & 96.90\% & 96.04\% & 95.60\% & 94.68\% & 93.83\% & 93.33\% & 98.48\% \\
                           &                     & acc on $D_f$ & 99.79\%  & 0.00\%  & 0.00\%  & 7.72\%  & 8.54\%  & 5.14\%  & 0.00\%  & 0.00\%  & 0.00\%  & 0.00\%  & 0.00\%  \\ 
\cline{1-14}
\multirow{4}{*}{CIFAR-100} & \multirow{2}{*}{1}  & acc on $D_r$ & 99.11\%  & 99.10\% & 97.92\% & 77.23\% & 98.72\% & 98.41\% & 97.64\% & 97.91\% & 98.59\% & 98.15\% & 99.33\% \\
                           &                     & acc on $D_f$ & 100.00\% & 0.00\%  & 0.00\%  & 4.14\%  & 0.00\%  & 5.65\%  & 0.00\%  & 0.00\%  & 0.00\%  & 0.00\%  & 0.00\%  \\
                           & \multirow{2}{*}{20} & acc on $D_r$ & 99.09\%  & 99.02\% & 64.29\% & 68.17\% & 91.53\% & 62.70\% & 72.08\% & 67.79\% & 65.34\% & 67.89\% & 76.58\% \\
                           &                     & acc on $D_f$ & 99.07\%  & 0.00\%  & 0.00\%  & 3.77\%  & 10.26\% & 12.38\% & 4.59\%  & 1.36\%  & 1.63\%  & 0.57\%  & 0.62\%  \\ \hline
\end{tabular}
\label{acc on Df and Dr train}
\end{sidewaystable}

\begin{figure*}[ht]
\centering
    \subfloat[]{\includegraphics[width=0.33\textwidth]{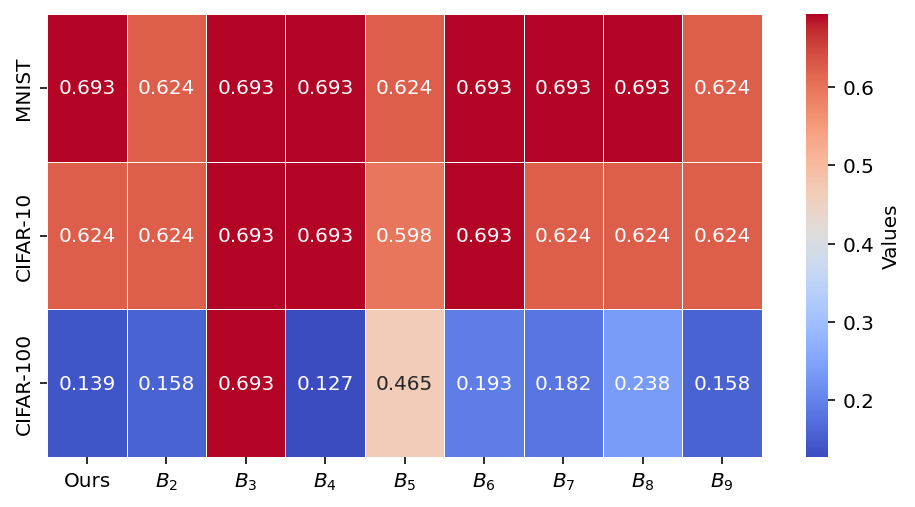}\label{trainsetmetrics-JSD}}
    \subfloat[]{\label{trainsetmetrics-L2}\includegraphics[width=0.33\textwidth]{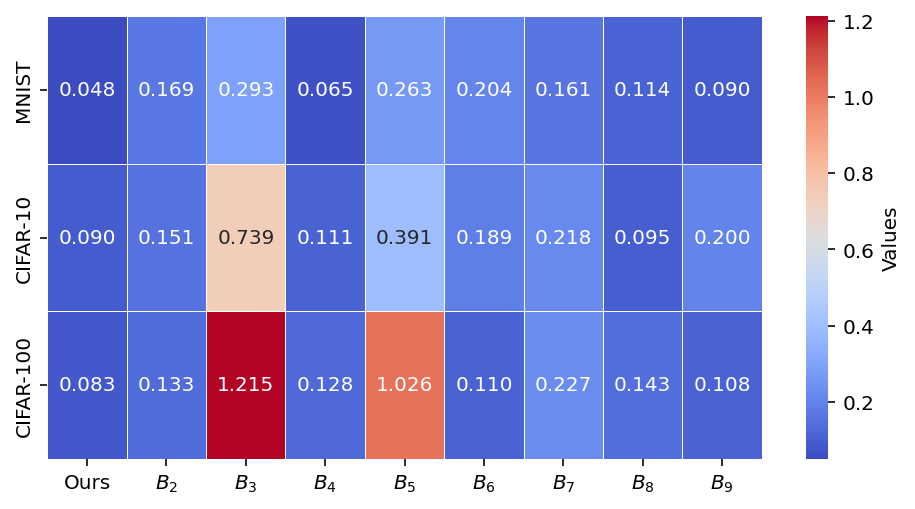}}
    \subfloat[]{\label{trainsetmetrics-pvalue}\includegraphics[width=0.33\textwidth]{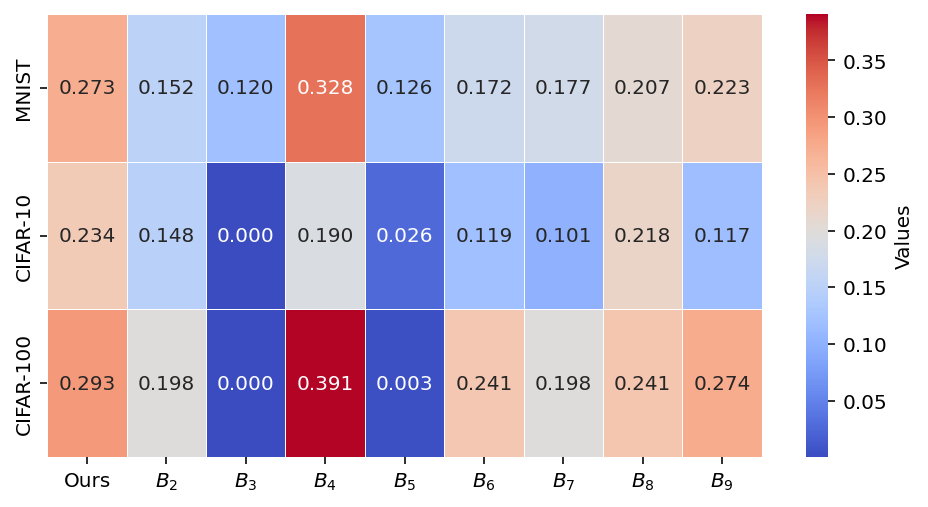}}
 \caption{Performance Metrics on Train Set. (a) JSD on Train Set, (b) L2 on Train Set, (c) p-value on Train Set.}
 \label{fig:JSD,l2,and p train}
\end{figure*} 




\end{appendices}}

\end{document}